\documentclass[12pt]{elsarticle}
\usepackage[utf8]{inputenc}
\usepackage{placeins}
\usepackage{graphicx}
\usepackage[a4paper,width=160mm,top=25mm,bottom=25mm,bindingoffset=6mm]{geometry}
\usepackage[compact]{titlesec}
\usepackage{hyperref}
\usepackage{tikz}
\usetikzlibrary{calc}
\usepackage{eso-pic}
\usepackage{setspace}
\usepackage{lipsum}
\usepackage[english]{babel}
\usepackage{fancyhdr}
\usepackage{array}
\usepackage{tabu}
\usepackage{multirow}
\usepackage{gensymb}
\usepackage{rotating}
\usepackage{caption}
\usepackage{subcaption}
\usepackage{hanging}
\usepackage{todonotes}
\usepackage{changepage}
\usepackage{subcaption}
\usepackage{lscape}
\usepackage{chngcntr}
\usepackage{float}
\floatstyle{plaintop}
\restylefloat{table}
\usepackage{xcolor}
\usepackage{amsmath}
\usepackage{mathtools}
\hypersetup{
    colorlinks,
    linkcolor={blue},
    citecolor={blue},
    urlcolor={blue}
}
\usepackage{ragged2e}
\usepackage{tikz, blindtext}
\usepackage{fix-cm}
\usepackage{enumitem}
\usepackage{multicol}
\usepackage{longtable}
\usepackage{amsfonts}
\usepackage[font=footnotesize]{caption}

\usepackage[english]{babel}

\newcommand{\jump}[1]{\ensuremath{[\![#1]\!]}}

\begin{document}

\begin{frontmatter}
    
\title{A discontinuous Galerkin / cohesive zone model approach\\ 
for the computational modeling of fracture\\ 
in geometrically exact slender beams}

\author[a,c]{Sai Kubair Kota}

\author[b]{Siddhant Kumar}

\author[a,c]{Bianca Giovanardi\corref{corauthor}}
\ead{B.Giovanardi@tudelft.nl}

\affiliation[a]{organization={Faculty of Aerospace Engineering},
addressline={Delft University of Technology}, 
state={2629 HS Delft},
country={The Netherlands}}

\affiliation[b]{organization={Department of Materials Science and Engineering},
addressline={Delft University of Technology}, 
state={2628 CD Delft},
country={The Netherlands}}

\affiliation[c]{organization={Delft Institute for Computational Science and Engineering},
addressline={Delft University of Technology}, 
state={2628 CD Delft},
country={The Netherlands}}

\cortext[corauthor]{Corresponding author}

\begin{abstract}
Slender beams are often employed as constituents in engineering materials and structures.
Prior experiments on lattices of slender beams have highlighted their complex failure response, 
where the interplay between buckling and fracture plays a critical role.
In this paper, we introduce a novel computational approach for modeling fracture in slender beams 
subjected to large deformations. We adopt a state-of-the-art geometrically exact Kirchhoff beam 
formulation to describe the finite deformations of beams in three-dimensions.
We develop a discontinuous Galerkin finite element discretization of the beam governing equations, 
incorporating discontinuities in the position and tangent degrees of freedom at 
the inter-element boundaries of the finite elements. Before fracture initiation, we enforce 
compatibility of nodal positions and tangents weakly, via the exchange of 
variationally-consistent forces and moments at the interfaces between adjacent elements.
At the onset of fracture, these forces and moments transition to cohesive laws modeling 
interface failure. We conduct a series of numerical tests to verify our computational framework
against a set of benchmarks and we demonstrate its ability to capture the tensile and bending 
fracture modes in beams exhibiting large deformations. 
Finally, we present the validation of our framework against fracture experiments of dry spaghetti 
rods subjected to sudden relaxation of curvature.
\end{abstract}




\end{frontmatter}

\pagestyle{plain}

\section{Introduction}

Slender beams are essential constituents of engineering materials and have a long history of serving 
as reinforcement elements in composite laminates~\cite{sharma2023critical}, 
textiles~\cite{grishanov2011structure}, and paper products~\cite{simon2021review}.
More recently, developments in additive manufacturing technology have enabled the combination of slender 
beams into engineered periodic truss lattices, giving rise to \emph{truss architected materials} 
\cite{deshpande2001foam,evans2001topological}. 
In the wake of these technological advancements, there has been growing emphasis on designing optimal 
material microstructures capable of yielding specific target mechanical properties. 
For example, recent efforts within the scientific community have focused on designing material architectures 
aimed at achieving a desired anisotropic stiffness \cite{bastek2022inverting}, optimal vibration control 
\cite{li2014active,guo2020dynamic}, high specific stiffness and strength 
\cite{meza2015resilient,abueidda2016effective,tancogne20183d}, and unprecedented specific impact energy absorption 
\cite{guell2019ultrahigh,portela2021supersonic}.
While characterizing the failure modes of these materials is of paramount importance, particularly in the context of 
the performance-weight trade-off, the fracture mechanics of truss architected materials is not completely understood 
and remains an area of ongoing scientific interest.

Compression experiments on truss architected materials have highlighted the relevance of the 
interplay of buckling and fracture of the individual beam structural constituents in their overall 
failure response \cite{meza2015resilient}.
While buckling has been extensively studied in the literature in isolation, the interplay between buckling and fracture
has traditionally received limited attention, perhaps due to the fact that buckling often leads to structural 
failure well before fracture initiation.
More recently, as the structural engineering community has broadened its focus from 
developing buckling-safe structures to leveraging buckling instabilities as a design opportunity to 
create structures capable of adapting their shape to their surrounding 
environment \cite{kochmann2017exploiting,vangelatos2019architected,lu2022architectural}, 
the study of the complex interplay between buckling and fracture in slender structures 
has gained increasing relevance.

Clearly, the complex fracture behavior of truss architected materials is significantly 
influenced by the combination of responses of the individual beam constituents. 
However, it is noteworthy that the buckling-to-fracture transition in a single slender beam already 
exhibits remarkable richness and complexity.
A celebrated example is the fracture behavior of dry spaghetti rods, which consistently break into 
more than two pieces when subjected to large pure-bending stresses.
This intriguing fracture behavior has puzzled numerous scientists, including the great physicist 
Richard Feynman \cite{sykes1996}.
Audoly and Neukirch later shed light on this phenomenon by uncovering a peculiar mechanical behavior 
of elastic rods in that the removal of stress leads to an \emph{increase} in strain 
\cite{audoly2005fragmentation}.
More specifically, the authors of that study theoretically predicted and confirmed through extensive
experimentation that the sudden relaxation of curvature can trigger a burst of flexural
waves, which locally increase the rod's curvature, ultimately leading to cascading fragmentation.

While physical experiments are essential for understanding the fracture mechanics of slender beams, 
computational approaches offer complementary insights, especially in scenarios where experimental 
methods become impractical or where the efficient exploration of extensive parameter spaces is 
necessary.
Preliminary research efforts aimed at developing computational models for fracture in beams 
can be found in the works of Armero and Ehrlich \cite{armero2006numerical}, Becker and Noels 
\cite{becker2011fracture} and, more recently, Lai et al. 
\cite{lai2020phase}. 
These studies are based on Euler-Bernoulli beam theory \cite{ochsner2021classical} but 
propose different approaches for 
fracture mechanics.
Armero and Ehrlich \cite{armero2006numerical} describe material failure via softening hinges 
modeled with the strong discontinuity approach. This approach, pioneered by Simo et al. 
\cite{simo1993analysis}, introduces discontinuities in the form of jumps of the solution field, 
which allow the characterization of the localized dissipative mechanisms associated with strain 
softening.
Becker and Noels \cite{becker2011fracture} adopt the discontinuous Galerkin / cohesive zone 
model approach, originally introduced by Radovitzky et al. \cite{radovitzky2011scalable}.
This computational fracture mechanics approach employs a discontinuous Galerkin 
discretization of the governing equations and models fracture as a process of
decohesion across interfaces between finite elements. 
In contrast, Lai et al. \cite{lai2020phase} describe the fracture process via a phase-field model 
\cite{francfort:1998,bourdin:2000}. 
Although these computational models have been successful in modeling failure in beams exhibiting 
small deformations, they cannot be used to model buckling-to-fracture transition, 
due to the infinitesimal-deformations assumption inherent in the underlying Euler-Bernoulli 
beam theory \cite{ochsner2021classical}. Likewise, beam fracture models that rely 
on Timoshenko beam theory \cite{ochsner2021classical}, 
e.g. \cite{ehrlich2005finite,bitar2018generalized,tojaga2021modeling},
are also unsuitable for capturing the transition from buckling to fracture.

Clearly, a fundamental requirement for a computational model for fracture of beams experiencing 
significant deformations is its foundation in a beam model able to accurately describe geometric 
nonlinearities 
\footnote{See the introduction of Meier et al. \cite{meier2019geometrically} for a recent 
review of geometrically exact beam models.}.
For example, Heisser et al. \cite{heisser2018controlling} found their beam fracture model on 
Kirchhoff beam theory, while Tojaga et al. \cite{tojaga2023geometrically} model fracture based on 
the finite-strain beam formulation by Simo \cite{simo1985finite}.
More specifically, Heisser et al. \cite{heisser2018controlling} model the fragmentation of a beam by 
disconnecting adjacent elements instantaneously, upon satisfaction of a stress-based fracture 
criterion. However, this approach neglects the time-dependent aspects of the fracture process, which
are known to be significant, particularly in the context of dynamic fragmentation 
\cite{miller1999modeling,zhou2006effects}.
By contrast, Tojaga et al. \cite{tojaga2023geometrically} employ the strong discontinuity approach 
discussed above. 
In their work, the authors enrich the displacement field by introducing discontinuous modes at the 
elements midpoints. Beyond a critical load, they model failure at these discontinuities as a 
softening hinge. 
It is important to highlight that their approach models discontinuities exclusively in the displacement 
degrees of freedom, but not in the rotation degrees of freedom. As a consequence, their framework is 
capable of capturing fracture modes arising from tension and shear but not those resulting from 
bending.

In this paper, we present a computational framework to model the tensile and bending modes of fracture in 
slender beams subjected to finite deformation. We model the deformation of the beam with the 
geometrically exact torsion-free Kirchhoff beam finite element framework by Meier et al. 
\cite{meier2014objective,meier2015locking} and we adopt the discontinuous Galerkin / cohesive zone 
model (DG/CZM) approach for fracture mechanics. Specifically, we approximate the beam's 
deformation based on the finite element method using third-order Hermitian polynomial shape functions and mixed interpolation 
involving both position and tangent degrees of freedom. This choice is well-suited for the spatial discretization 
of the Kirchhoff beam formulation in view of the ability of third-order Hermitian polynomials 
to meet the $C^1$ continuity requirement. However, instead of imposing the compatibility strongly, we adopt a 
discontinuous Galerkin finite element approach and incorporate discontinuities in the position and tangent degrees of 
freedom at the inter-element boundaries of the finite elements. Before fracture initiation, we enforce 
compatibility of nodal positions and tangents weakly, via the exchange of 
variationally-consistent forces and moments at the interfaces between adjacent elements.
At the onset of fracture, these variationally-consistent forces and moments transition 
to cohesive laws that model the fracture process.
The finite element discretization described above results in a time-dependent algebraic system, 
which is solved in time with the second-order explicit Newmark scheme.

The resulting computational framework effectively captures both tensile and bending modes of 
fracture of slender beams in the geometrically nonlinear regime. Compared to existing 
state-of-the-art computational models for beam fracture, our approach offers the following key 
advantages.
Unlike the model proposed by Heisser et al. \cite{heisser2018controlling}, which lacks an energy 
dissipation mechanism, our approach is firmly rooted in a sound physical model for fracture 
mechanics.
Consequently, it is able to describe the energy dissipation resulting from the fracture 
process, eliminating the need for ad-hoc measures such as enforcing a minimal fragment length to 
prevent unphysical crack formations \cite{heisser2018controlling}.
Additionally, unlike the approach presented by Tojaga et al. \cite{tojaga2023geometrically}, our 
approach is able to describe the bending fracture modes.

We conduct a series of numerical tests to verify our computational framework against a set of 
benchmarks and demonstrate its capability of accurately modeling tensile and bending fracture in 
slender beams exhibiting large deformations. 
First, we verify that the discontinuous Galerkin discretization is able to capture the analytical 
buckling load for a column. We then verify the discontinuous Galerkin / cohesive zone model 
fracture mechanics approach in a bar spall fracture benchmark. Next, we show that the incorporation 
of discontinuities in the tangent degrees of freedom is essential for capturing the bending 
mode of fracture. Finally, we apply our computational framework to reproduce experiments by Audoly 
and Neukirch \cite{audoly2005fragmentation} on the fracture of dry spaghetti rods bent and suddenly 
released.

The structure of the paper is as follows. In Section \ref{sec:governing-equations}, we briefly review 
the geometrically exact Kirchhoff beam formulation of \cite{simo1985finite,meier2014objective}. 
Section \ref{sec:cohesive-zone-modeling} discusses the resultant based cohesive zone modeling approach for tensile 
and bending fracture of beams. In Section \ref{sec:dg-formulation}, we derive the discontinuous Galerkin weak 
formulation of the beam governing equations and the DG/CZM weak formulation 
to model the tensile and bending modes of fracture in slender beams.
We, then, outline the space and time discretization of the DG/CZM weak form and discuss our solution 
strategy for the discrete system of equations.
We perform thorough verification and validation of the computational framework in Section 
\ref{sec:results}.
Conclusions are drawn in Section \ref{sec:conclusions}.

\pagestyle{plain}

\section{Geometrically exact Kirchhoff beam formulation} \label{sec:governing-equations}

For completeness, we provide a concise summary of the geometrically exact beam governing equations 
by Simo~\cite{simo1985finite} in their shear-free variant, as derived by Meier et 
al.~\cite{meier2014objective}. We refer the reader to those two works for a more detailed and
comprehensive discussion.

\subsection{Kinematics}

\begin{figure}[h!]
	\begin{center}
		\includegraphics[width=0.8\textwidth]{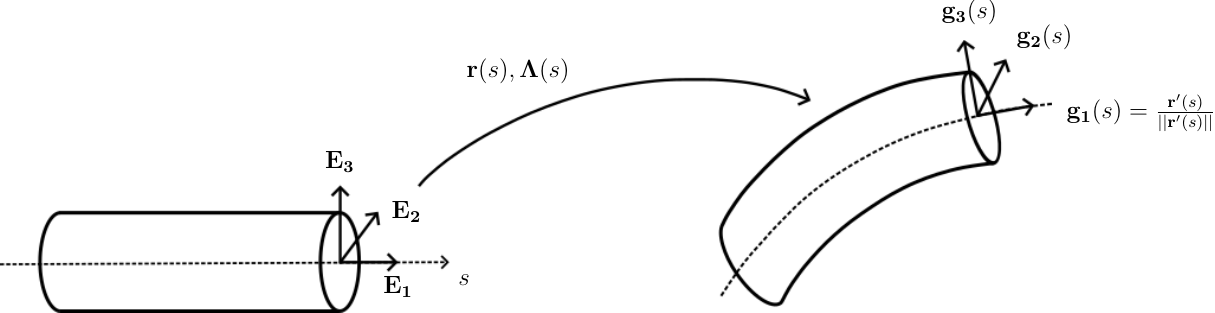}
	\end{center}
	\caption{\textit{Illustration of the kinematics of a geometrically exact Kirchhoff beam in the case of 
    a beam with circular cross-section. The beam configuration is 
    characterized by the line of centroids $\boldsymbol{r}(s)$ and by the orthonormal 
    \emph{intrinsic} frame $\{\boldsymbol{g_1}(s), \boldsymbol{g_2}(s), \boldsymbol{g_3}(s)\}$. 
    The Kirchhoff constraint enforces that $\boldsymbol{g_1}(s)$ be tangent to the line of 
    centroids, that is 
    $\boldsymbol{g_1}(s) = \frac{\boldsymbol{r}^{\prime}(s)}{||\boldsymbol{r}^{\prime}(s)||}$.}}
	\label{fig:beam-kinematics}
\end{figure}

Following Simo~\cite{simo1985finite}, we characterize the beam configuration by the position of the 
beam centerline and by the orientation of its cross-sections.
The centerline of the beam, i.e. the curve of the cross-sections centroids, is described with 
a suitable parametrization $\boldsymbol{r}(s) \in \mathbb{R}^3$, where $s \in [0,L]$ 
is the arc-length parameter, while the orientation of the beam cross-sections is described 
in terms of the orthonormal \emph{intrinsic} frame $\{\boldsymbol{g_1}(s), \boldsymbol{g_2}(s), 
\boldsymbol{g_3}(s)\}$, see Figure \ref{fig:beam-kinematics}.
By convention, $\boldsymbol{g_1}(s)$ is chosen orthogonal to the beam cross-section at 
$s$, while $\boldsymbol{g_2}(s)$ and $\boldsymbol{g_3}(s)$ are chosen parallel to 
its principal axes of inertia. Note that, in general, $\boldsymbol{g_1}(s)$ is not tangent to the 
line of centroids $\boldsymbol{r}(s)$.

The deformed configuration of the beam can be expressed in terms of $\boldsymbol{r}$, 
$\boldsymbol{g_2}$, and $\boldsymbol{g_3}$ as:
\begin{equation} \label{eq:deformed-configuration}
\boldsymbol{x}(s, \xi_2, \xi_3) = \boldsymbol{r}(s) + \xi_2 \ \boldsymbol{g_2}(s) 
    + \xi_3 \ \boldsymbol{g_3}(s),
\end{equation}
where $\xi_2$ and $\xi_3$ are coordinates on the beam cross-section in the reference configuration.

For simplicity, it is often assumed that the beam is straight in its reference configuration, so 
that the intrinsic frame in the reference configuration can be chosen as a fixed basis
$\{\boldsymbol{E_1},\boldsymbol{E_2},\boldsymbol{E_3}\}$ of $\mathbb{R}^3$. 
\footnote{This assumption is not fundamental and can be removed as discussed in 
\cite{simo1993analysis}.}
Because the vector basis $\{\boldsymbol{g_i}(s)\}$ is orthonormal 
at any cross-section $s$, there exists a rotation tensor $\boldsymbol{\Lambda}(s)$ such that:
\begin{equation} \label{eq:gi-rotation}
\boldsymbol{g_i}(s) = \boldsymbol{\Lambda}(s) \boldsymbol{E_i}, \quad i = 1, 2, 3.
\end{equation}

Kinematically admissible variations of $\boldsymbol{r}$ and $\boldsymbol{g_i}$ are denoted with 
$\boldsymbol{\delta r}$ and $\boldsymbol{\delta g_i}$, respectively.
A direct consequence of Equation \eqref{eq:gi-rotation} is that kinematically admissible variations of 
$\boldsymbol{g_i}$ can be expressed as
\begin{equation} \label{eq:delta-gi-skew-symm}
\boldsymbol{\delta g_i}(s) = \boldsymbol{\delta \Lambda}(s)\boldsymbol{\Lambda^T}(s) \boldsymbol{g_i}(s).
\end{equation}
Because tensor $\left(\boldsymbol{\delta \Lambda}\right)\boldsymbol{\Lambda^T}$ is skew-symmetric,
Equation \eqref{eq:delta-gi-skew-symm} can be rewritten as
\begin{equation} \label{eq:delta-gi}
\boldsymbol{\delta g_i}(s) = \boldsymbol{\delta \theta}(s) \times \boldsymbol{g_i}(s),
\end{equation}
where $\boldsymbol{\delta \theta}$ is the axial vector of 
$\left(\boldsymbol{\delta \Lambda}\right)\boldsymbol{\Lambda^T}$ 
and is also referred to as the \emph{spin vector}.
Under the hypothesis of negligible shear strains, which is a well-accepted assumption in the case
of slender beams \cite{meier2014objective}, $\boldsymbol{\delta \theta}$ is not independent of 
$\boldsymbol{\delta r}^{\prime}$. In fact, a relation between $\boldsymbol{\delta \theta}$ and 
$\boldsymbol{\delta r}^{\prime}$ can be obtained by kinematically enforcing that the 
beam centerline remain perpendicular to the cross-sections during the deformation 
(\emph{Kirchhoff constraint}):
\begin{equation} \label{eq:kirchhoff-constraint}
\boldsymbol{g_1}(s) = \frac{\boldsymbol{r}^\prime(s)}{||\boldsymbol{r}^\prime(s)||},
\end{equation}
where $(\cdot)^{\prime}$ denotes the derivative with respect to the arc-length parameter.
Plugging Equation \eqref{eq:kirchhoff-constraint} into Equation \eqref{eq:delta-gi}, we 
obtain the following relation:
$$
\boldsymbol{\delta\theta}(s) = 
    \frac{\boldsymbol{r}^{\prime}(s) \times 
    \boldsymbol{\delta r}^{\prime}(s)}{||\boldsymbol{r}^{\prime}(s)||^2}
    + \frac{\boldsymbol{r}^{\prime}(s)}{||\boldsymbol{r}^{\prime}(s)||}\delta \alpha,
$$
prescribing the kinematically admissible variations of rotations $\boldsymbol{\delta\theta}$
in terms of the kinematically admissible variations of the centerline tangents 
$\boldsymbol{\delta r}^{\prime}$ and the tangential component of the spin vector:
$$
\delta \alpha(s) := \boldsymbol{\delta \theta}(s) \cdot 
    \frac{\boldsymbol{r}^{\prime}(s)}{||\boldsymbol{r}^{\prime}(s)||}.
$$
where $\delta\alpha$ represents the kinematically admissible variation of the total twist angle.
In this work, we limit our attention to the simpler torsion-free formulation
and, following \cite{meier2015locking}, we completely resign the degrees of freedom representing the 
torsional deformation modes, resulting in:
\begin{equation} \label{eq:delta-theta-r}
    \boldsymbol{\delta\theta}(s) =\boldsymbol{\delta\theta}_{\perp}(s) := 
        \frac{\boldsymbol{r}^{\prime}(s) \times 
        \boldsymbol{\delta r}^{\prime}(s)}{||\boldsymbol{r}^{\prime}(s)||^2}.
\end{equation}

We refer the reader to \cite{meier:2016} for a comprehensive discussion on the conditions under 
which a torsion-free beam formulation is appropriate.

\subsection{Balance of linear and angular momentum}

The governing equations of the beam can be derived by integrating the linear and angular momentum 
balance equations from the 3D continuum theory over the cross-section of the beam. We refer the 
reader to Simo~\cite{simo1985finite} for a detailed derivation, while we report here only the 
final expressions:

\begin{gather}
	\boldsymbol{f}^{\prime} + \boldsymbol{\tilde{f}} = \rho A\boldsymbol{\ddot{r}},\label{eq1} \\
	\boldsymbol{m}^{\prime} + \boldsymbol{r}^{\prime} \times \boldsymbol{f} + \boldsymbol{\tilde{m}} = \boldsymbol{I_\rho} \boldsymbol{\dot{\omega}} + \boldsymbol{\omega\times (I_\rho\omega)}.\label{eq2}
\end{gather}

\noindent
Here, the prime and dot symbols denote arc-length and material time derivatives, respectively,
while $\rho$, $A$, and $\boldsymbol{I_\rho}$ are the referential mass density, area of 
cross-section, and spatial inertia tensor of the beam.
$\boldsymbol{\omega}$ is the axial vector corresponding to the the skew-symmetric 
angular velocity tensor $\dot{\boldsymbol{\Lambda}}\boldsymbol{\Lambda}^{T}$, while 
$\boldsymbol{f}(s, t)$ and $\boldsymbol{m}(s, t)$ are the internal force and moment stress 
resultants:

\begin{gather}
	\boldsymbol{f} = \int_{A} \boldsymbol{P}\boldsymbol{E_1} \ \text{d}A, \\
	\boldsymbol{m} = \int_{A} \left(\boldsymbol{x} - \boldsymbol{r}\right) 
        \times \boldsymbol{P}\boldsymbol{E_1} \ \text{d}A,
\end{gather}
where $\boldsymbol{P}$ is the Piola-Kirchhoff stress tensor.
Finally, $\boldsymbol{\tilde{f}}$ and $\boldsymbol{\tilde{m}}$ are the external distributed forces 
and moments per unit referential arc-length.

Forces $\boldsymbol{f}$, $\boldsymbol{\tilde{f}}$ and moments $\boldsymbol{m}$, 
$\boldsymbol{\tilde{m}}$ can be additively decomposed into axial $\boldsymbol{f_{||}}$, 
$\boldsymbol{\tilde{f}_{||}}$ and shear $\boldsymbol{f_{\perp}}$, $\boldsymbol{\tilde{f}_{\perp}}$ 
forces, and bending $\boldsymbol{m_{\perp}}$, $\boldsymbol{\tilde{m}_{\perp}}$ and torsional 
$\boldsymbol{m_{||}}$, $\boldsymbol{\tilde{m}_{||}}$ moments, where the 
notation~\eqref{eq:vector-decomposition} is used.

\begin{equation}
    \boldsymbol{a_{||}} := \left( \boldsymbol{a} 
    \cdot \frac{\boldsymbol{r}^{\prime}}{||\boldsymbol{r}^{\prime}||} \right) 
    \frac{\boldsymbol{r}^{\prime}}{||\boldsymbol{r}^{\prime}||},
    \qquad \boldsymbol{a_{\perp}} := \boldsymbol{a} - \boldsymbol{a_{||}}. \label{eq:vector-decomposition}
\end{equation}

Under the assumption that rotational inertia (i.e. the right-hand side of Equation \eqref{eq2}) can 
be neglected, which is a well-accepted assumption for the case of slender beams 
\cite{boyer2004,meier2016finite}, the following equation expressing the internal shear forces as a 
function of the bending moments can be obtained by performing the cross product of Equation 
\eqref{eq2} with $\boldsymbol{r}^{\prime}$:
\begin{equation}\label{eq:shear-forces}
  \boldsymbol{f_{\perp}} = \frac{\boldsymbol{r}^{\prime}}{||\boldsymbol{r}^{\prime}||^{2}} \times \big(\boldsymbol{m_{\perp}}^{\prime} + \boldsymbol{\tilde{m}_{\perp}}\big)
\end{equation}
Equation \eqref{eq:shear-forces} can be plugged into Equation \eqref{eq1}, leading to:
\begin{equation}\label{eq3}
	\boldsymbol{f_{||}}^{\prime} + \Bigg[\frac{\boldsymbol{r}^{\prime}}{||\boldsymbol{r}^{\prime}||^{2}} \times \big(\boldsymbol{m^{\prime}_{\perp}} + \boldsymbol{\tilde{m}_{\perp}}\big)\Bigg]^{\prime} + \boldsymbol{\tilde{f}} = \rho A\boldsymbol{\ddot{r}}.
\end{equation}

Note that, in the torsion-free formulation, Equation \eqref{eq:shear-forces} is equivalent to Equation 
\eqref{eq2}. As a consequence, Equation \eqref{eq3} is equivalent to the original system of 
governing equations \eqref{eq1} and \eqref{eq2}. In this work, we focus on the simpler 
torsion-free case and we use Equation \eqref{eq3} to model the finite deformations 
of the beam.

Equation~\eqref{eq3} is complemented with suitable initial conditions for positions and tangents,
and boundary conditions, in terms of applied forces and bending moments on the Neumann boundary 
or imposed positions and tangents on the Dirichlet boundary.

\subsection{Constitutive equations}

\noindent
In this study, we confine our attention to isotropic beams with circular cross-sections.
Assuming hyperelastic material behavior, the internal axial forces and bending moments are related 
to the axial strain $\epsilon = ||\boldsymbol{r}^{\prime}|| - 1$ and the curvature 
$\boldsymbol{\kappa} = \frac{\boldsymbol{r}^{\prime} \times 
    \boldsymbol{r^{\prime\prime}}}{||\boldsymbol{r}^{\prime}||^2}$
through the following constitutive relations \cite{meier2015locking}:
\begin{equation}\label{eq:constitutive}
	\boldsymbol{f_{||}} = EA\epsilon\frac{\boldsymbol{r}^{\prime}}{||\boldsymbol{r}^{\prime}||}, \qquad
	\boldsymbol{m_{\perp}} = EI\boldsymbol{\kappa},
\end{equation}
where $E$, $A$ and $I$ are the Young's modulus, cross-sectional area and moment of 
inertia of the beam, respectively.

\noindent
It should be emphasized that the constitutive relations~\eqref{eq:constitutive} are only 
applicable in the context of small strains. 
Nevertheless, the proposed computational framework is general in the sense that the constitutive 
relations can be extended to incorporate various material behaviors. For instance, 
plasticity can be included to simulate fracture in metals, while viscoelasticity~\cite{ferri2023efficient} or 
elasto-visco-plasticity~\cite{weeger2022mixed} can be integrated to model fracture in 3D-printed polymers.

\section{Cohesive zone approach for fracture in beams}\label{sec:cohesive-zone-modeling}

\begin{figure}[h!]
	\begin{center}
		\includegraphics[width=0.80\textwidth]{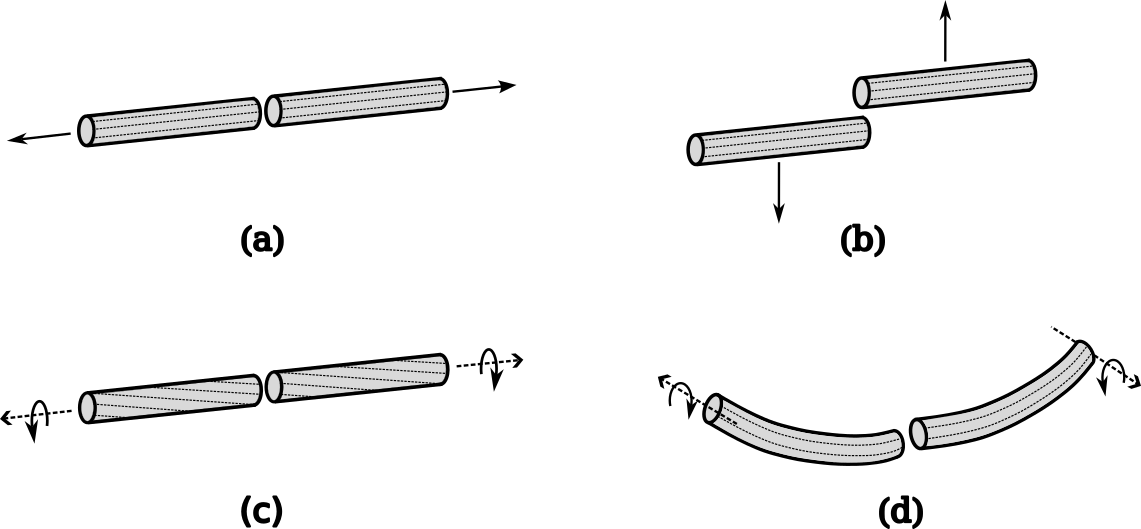}
	\end{center}
	\caption{\textit{The fracture modes of beams: (a) tension, (b) shear, (c) torsion 
    and (d) bending.}}
	\label{fig2_1}
\end{figure}

We discuss the fracture modes of beams and propose cohesive laws to model the tensile and bending modes of beam 
fracture in this section. 
Beams exhibit several modes of fracture, which we illustrate in Figure~\ref{fig2_1}.
In fact, beams can fail under tensile, transverse, torsional or bending loads. 
In this work, we 
focus our attention on the tensile and bending fracture modes in slender beams, 
while we neglect the shear and torsional fracture modes.

We adopt the cohesive zone approach to model the beam fracture behavior. Cohesive zone models employ a traction-separation 
law to characterize the evolution of a crack, under the assumption that fracture processes occur within a region of finite-length 
ahead of the crack tip, referred to as the cohesive zone. However, instead of 
employing the conventional traction-separation laws typical in cohesive zone models~\cite{camacho1996computational},
we formulate the cohesive laws in the stress resultant form, see~\cite{talamini2017parallel,zavattieri2006modeling}.
Following common practice in the cohesive zone modeling framework, we make the customary assumption that cohesive forces and moments
depend solely on the kinematic jumps in the proximity of the crack tip.

We introduce cohesive boundaries at interfaces of adjacent elements of the beam, 
where jump discontinuities in the kinematic fields can occur. At these interfaces we define the 
axial and bending kinematic jumps $\Delta_{||}$ and $\boldsymbol{\Theta}$ as:
\begin{equation}\label{eq:kinematic-jumps}
    \begin{aligned}
        \Delta_{||} &= \jump{\boldsymbol{r}}\cdot\boldsymbol{\hat{n}_{coh}} \\
	    \boldsymbol{\Theta} &= \jump{\boldsymbol{g_1}},
    \end{aligned}
\end{equation}
where $\boldsymbol{\hat{n}_{coh}}$ is the unit normal to the cohesive boundary in the current configuration:
\begin{equation}\label{eq14}
	\boldsymbol{\hat{n}_{coh}} = \frac{\langle \boldsymbol{g_1} \rangle}{||\langle\boldsymbol{g_1}\rangle||}.
\end{equation}
and the notations

\vspace{-0.5cm}
$$
\jump{\bullet}\Big|_{s = a} := \lim_{s\rightarrow a^+} \bullet - \lim_{s\rightarrow a^-} \bullet
$$
$$
\langle{\bullet}\rangle\Big|_{s = a} := \frac{1}{2}\left(\lim_{s\rightarrow a^+} \bullet + \lim_{s\rightarrow a^-} \bullet \right)
$$
\noindent
represent the jump and average of the field $\bullet$ at any arbitrary point $s = a \in (0, L)$.

\noindent
Note that $\boldsymbol{\Theta}$ has a vanishing component along the normal to the cohesive 
boundary, as $\boldsymbol{\Theta} \cdot \boldsymbol{\hat{n}_{coh}} = 0$ holds by construction.

To enable mixed-mode fracture under tension and bending, we introduce a scalar effective 
separation $\Delta$:
\begin{equation}
	\Delta = \sqrt[]{\{\Delta_{||}\}^2 + \big(\alpha R ||\boldsymbol{\Theta}|| \big)^2},
\end{equation}
where $R$ is the radius of the beam and $\alpha$ is a mode-mixity parameter, akin to that 
traditionally employed for mixed-mode cohesive laws \cite{ortiz1999}. 
In the equation above, $\{\boldsymbol{\cdot}\} = \max\:(\boldsymbol{\cdot}\:,\:0)$ denotes the 
Macaulay operator.

\begin{figure}[h!]
    \begin{center}
		\includegraphics[width=0.35\textwidth]{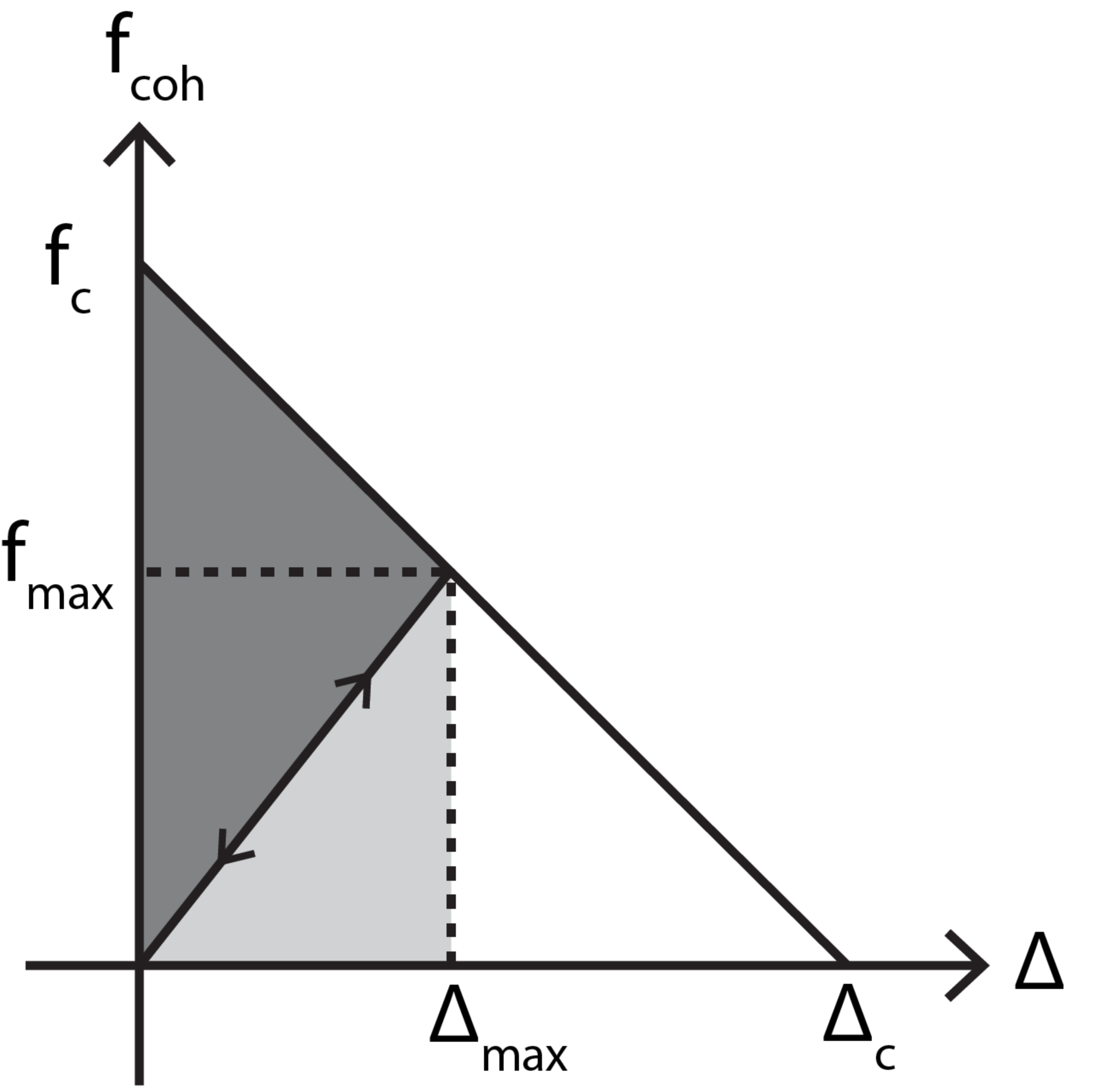}
	\end{center}
	\caption{\textit{The force-separation cohesive law prescribes a linear decay of the scalar 
    effective cohesive force $f_{coh}$ with the scalar effective separation $\Delta$, from a 
    critical value $f_c$ to zero. Irreversibility is modeled by introducing a history 
    variable $\Delta_{max}$ representing the maximum effective separation achieved. 
    The unloading path follows a trajectory back to the origin, while reloading occurs on the same 
    unloading path. The total area under the force-separation curve is equal to the effective 
    fracture energy of the material while the black and grey areas at any point $\Delta = \Delta_{max}$ 
    represent the dissipated and maximum recoverable energies at the cohesive boundary. Complete 
    fracture is achieved when $\Delta \geq \Delta_{c}$.}}
	\label{fig3}
\end{figure}

We introduce the following cohesive axial forces and cohesive bending moments resisting the opening 
of cracks in the beam:
\begin{equation}\label{eq:cohesive-forces-moments}
    \begin{aligned}
	\boldsymbol{f_{coh,\:||}} &= f_{coh}(\Delta,\boldsymbol{q})\:\frac{\{\Delta_{||}\}}{\Delta}\:\boldsymbol{\hat{n}_{coh}}, \\
	\boldsymbol{m_{coh,\:\perp}} &= \alpha^2\:f_{coh}(\Delta,\boldsymbol{q})\:\frac{R^2}{\Delta}\:\boldsymbol{\Theta},
    \end{aligned}
\end{equation}
where the effective cohesive force $f_{coh}$ is a scalar function of the effective separation 
$\Delta$ and of a set of internal variables $\boldsymbol{q}$. 
The function $f_{coh}(\Delta,\boldsymbol{q})$ can be tailored based on the desired representation of the constitutive 
fracture behavior (e.g. brittle, quasi-brittle, ductile).
Here, we assume that cohesive axial forces and cohesive bending moments decay linearly with $\Delta$, 
as illustrated in Figure~\ref{fig3}, and account for irreversibility by introducing 
a history internal variable $\Delta_{max}$, as is customary in cohesive zone modeling.

Specifically, in the loading stage, we set:
\begin{equation}
	f_{coh}(\Delta,\Delta_{max}) = \Bigg(1-\frac{\Delta}{\Delta_c}\Bigg)\:f_c\quad\text{for}\quad\Delta \geq \Delta_{max},
\end{equation}
where $\Delta_{max}$ is the maximum effective separation in the entire loading history, 
$\Delta_c = 2G_c / \sigma_c$ is the effective separation at which complete decohesion occurs, and 
$G_c$ is the effective fracture energy.
In the unloading and reloading stage, we set:
\begin{equation}
	f_{coh}(\Delta,\Delta_{max}) = \frac{\Delta}{\Delta_{max}}f_{max}\quad\text{for}\quad\Delta < \Delta_{max},
\end{equation}
where $f_{max}$ is the effective cohesive force at $\Delta_{max}$.

The cohesive laws described above are activated at an inter-element boundary of the beam
upon meeting the following fracture initiation criterion:
\begin{equation}\label{eq18}
	f_{eq}(\langle\boldsymbol{f}\rangle,\langle\boldsymbol{m_{\perp}}\rangle) \geq f_{c},
\end{equation}
where $f_c = \sigma_c A$ is the critical effective cohesive force expressed in terms of the 
material's cohesive strength $\sigma_c$, and $f_{eq}$ is an equivalent force given by the scalar:
\begin{equation}\label{eq19}
	f_{eq}(\boldsymbol{f},\boldsymbol{m_{\perp}}) = \sqrt[]{\{\boldsymbol{f} \cdot \boldsymbol{\hat{n}_{coh}} \}^2 + \Bigg|\Bigg|\frac{\boldsymbol{m_{\perp}}}{\alpha R}\Bigg|\Bigg|^2}.
\end{equation}

\section{Computational framework for fracture in geometrically exact slender beams} 
\label{sec:dg-formulation}

In this section, we derive our discontinuous Galerkin / cohesive zone model approach for 
fracture in geometrically exact slender beams. We first derive the discontinuous Galerkin weak 
formulation of the beam governing equations presented in Section \ref{sec:governing-equations}. 
We, then, present the weak formulation of the discontinuous 
Galerkin / cohesive zone model followed by its space and time discretization.

\subsection{Derivation of the discontinuous Galerkin weak form} \label{subsec:dg-formulation}

\begin{figure}[h!]
	\vspace{0.5cm}
	\begin{center}
		\includegraphics[width=0.75\textwidth]{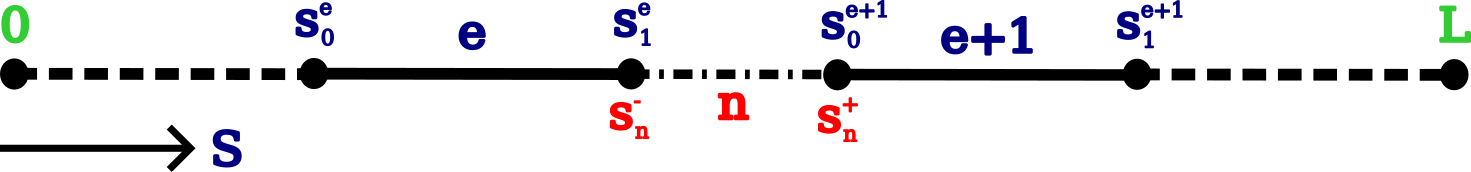}
	\end{center}
	\caption{\textit{The discontinuous Galerkin discretization of the straight undeformed beam. 
    Internal nodes, e.g. nodes $s_1^e$ and $s_0^{e+1}$, are duplicated to allow the embedding of 
    potential discontinuities at the element interfaces $s_n$ for $n = 1, ..., E-1$.}}
	\label{fig2}
\end{figure}

We consider a space discretization $\Omega_{h}$ of the straight undeformed beam $\Omega$ into 
segments $\Omega_{e} = (s_0^e, s_1^e)$, $e = 1, ..., E$, so that 
$\Omega_{h} = \bigcup\limits_{e=1}^{E} \overline{\Omega}_{e}$, see schematic in Figure~\ref{fig2}.
We denote with $\partial \Omega_{h} = \{0,L\}$ the boundary of the beam, whose external normal $n_e$ 
is $n_e = 1$ at $s=L$ and $n_e = -1$ at $s=0$. Finally, we denote with $\partial_{N_f}\Omega_h$ and 
$\partial_{N_m}\Omega_h$ the Neumann portions of $\partial \Omega_{h}$, where we apply forces 
$\boldsymbol{\bar{f}}$ and moments $\boldsymbol{\bar{m}}_{\perp}$, respectively.

We consider Equation~\eqref{eq3}, where we allow $\boldsymbol{r}(s)$ and its associated kinematically 
admissible variations $\boldsymbol{\delta r}(s)$ to exhibit discontinuities at the interfaces 
$s_n$, $n = 1, ..., E-1$ of adjacent elements.
We start the derivation of the discontinuous Galerkin weak form of Equation~\eqref{eq3} by multiplying it 
with $\boldsymbol{\delta r}$, integrating over the individual subdomains $\Omega_{e}$, $e = 1, ..., E$,
and performing the integration by parts:

\begin{equation}\label{eq7_1}
    \begin{aligned}
    - \sum_{e=1}^{E} \int_{\Omega_{e}} \boldsymbol{f_{||}}    \cdot \boldsymbol{\delta r}^{\prime} \:ds 
    &- \sum_{e=1}^{E} \int_{\Omega_{e}} \boldsymbol{f_{\perp}} \cdot \boldsymbol{\delta r}^{\prime} \:ds
    - \sum_{n=1}^{E-1}\:\jump{\boldsymbol{f} \cdot \boldsymbol{\delta r}}\Big|_{s_{n}}
	+ \left(\boldsymbol{\bar{f}} \cdot \boldsymbol{\delta r} \right)\Big|_{\partial_{N_f}\Omega_h} \\
	& + \sum_{e=1}^{E} \int_{\Omega_{e}} \boldsymbol{\tilde{f}} \cdot \boldsymbol{\delta r} \:ds 
    = \sum_{e=1}^{E} \int_{\Omega_{e}} \rho A \boldsymbol{\ddot{r}} \cdot \boldsymbol{\delta r} \:ds,
    \end{aligned}
\end{equation}
where we have applied the Neumann boundary condition $\boldsymbol{\bar{f}} = n_e \boldsymbol{f}$ on 
$\partial_{N_f}\Omega_h$.
In Equation \eqref{eq7_1}, the notation $\bullet \Big|_a$ means $\bullet$ evaluated at $s = a$.

\noindent
Using the identity $\jump{\boldsymbol{a}\cdot \boldsymbol{b}} = \langle\boldsymbol{a}\rangle \cdot
\jump{\boldsymbol{b}} + \jump{\boldsymbol{a}}\cdot\langle\boldsymbol{b}\rangle$, we can rewrite 
the jump term in Equation~\eqref{eq7_1} as:
\begin{equation}\label{eq7_1bis}
    \sum_{n=1}^{E-1}\:\jump{\boldsymbol{f} \cdot \boldsymbol{\delta r}}\Big|_{s_{n}}
    = \sum_{n=1}^{E-1}\:\Big[\langle\boldsymbol{f}\rangle \cdot \jump{\boldsymbol{\delta r}} \Big]\Big|_{s_{n}} +
	\sum_{n=1}^{E-1}\:\Big[\jump{\boldsymbol{f}}\cdot \langle\boldsymbol{\delta r}\rangle \Big]\Big|_{s_{n}}.
\end{equation}

\noindent
In addition, Equation \eqref{eq:shear-forces}, together with the vector identity
$\boldsymbol{a}\cdot(\boldsymbol{b}\times\boldsymbol{c}) = \boldsymbol{c}\cdot(\boldsymbol{a}\times\boldsymbol{b})$, 
allows us to rewrite the second term of Equation \eqref{eq7_1} as:
\begin{equation}\label{eq7_2}
    \begin{aligned}
- \sum_{e=1}^{E} \int_{\Omega_{e}} \boldsymbol{f_{\perp}} \cdot \boldsymbol{\delta r}^{\prime} \:ds
& = \sum_{e=1}^{E} \int_{\Omega_{e}} \left(\boldsymbol{m_{\perp}}^{\prime} + \boldsymbol{\tilde{m}_{\perp}}\right) \cdot 
\frac{\boldsymbol{r}^{\prime} \times \boldsymbol{\delta r}^{\prime}}{||\boldsymbol{r}^{\prime}||^2} \:ds \\
& = \sum_{e=1}^{E} \int_{\Omega_{e}} \left(\boldsymbol{m_{\perp}}^{\prime} + \boldsymbol{\tilde{m}_{\perp}}\right) \cdot 
\boldsymbol{\delta \theta}_{\perp}\:ds.
    \end{aligned}
\end{equation}
Note that the kinematically admissible variation $\boldsymbol{\delta \theta}_{\perp}$ of 
Equation \eqref{eq:delta-theta-r} arises naturally as the work conjugate of the bending moments.
In the following derivation, we will omit the subscript in $\boldsymbol{\delta \theta}_{\perp}$ and 
simply write $\boldsymbol{\delta \theta}$ for the sake of a lighter notation.
The right-hand side of Equation \eqref{eq7_2} can be, in turn, integrated by parts, leading to:
\begin{equation}\label{eq7_3}
    \begin{aligned}
    & \sum_{e=1}^{E} \int_{\Omega_{e}} \left(\boldsymbol{m_{\perp}}^{\prime} + \boldsymbol{\tilde{m}_{\perp}}\right) \cdot 
\boldsymbol{\delta \theta}\:ds
        = \sum_{e=1}^{E} \int_{\Omega_{e}} \boldsymbol{\tilde{m}_{\perp}} \cdot \boldsymbol{\delta \theta}\:ds
        - \sum_{e=1}^{E} \int_{\Omega_{e}} \boldsymbol{m_{\perp}} \cdot \boldsymbol{\delta \theta}^{\prime}\:ds \\
    & \qquad
     - \sum_{n=1}^{E-1}\:\Big[\langle\boldsymbol{m_{\perp}}\rangle \cdot \jump{\boldsymbol{\delta \theta}} \Big]\Big|_{s_{n}}
     - \sum_{n=1}^{E-1}\:\Big[\jump{\boldsymbol{m_{\perp}}}\cdot \langle\boldsymbol{\delta \theta}\rangle \Big]\Big|_{s_{n}}
     + \left(\boldsymbol{\bar{m}_{\perp}} \cdot \boldsymbol{\delta \theta} \right)\Big|_{\partial_{N_m}\Omega_h},
    \end{aligned}
\end{equation}
where we have applied the Neumann boundary condition $\boldsymbol{\bar{m}}_{\perp} = n_e \boldsymbol{m}_{\perp}$ on 
$\partial_{N_m}\Omega_h$ and 
used again the identity $\jump{\boldsymbol{a}\cdot \boldsymbol{b}} = \langle\boldsymbol{a}\rangle \cdot
\jump{\boldsymbol{b}} + \jump{\boldsymbol{a}}\cdot\langle\boldsymbol{b}\rangle$.

Gathering Equations \eqref{eq7_1}, \eqref{eq7_1bis}, \eqref{eq7_2}, and \eqref{eq7_3}, we obtain:
\begin{equation}\label{eq:dg-weak-form}
    \begin{aligned}
    & \sum_{e=1}^{E} \int_{\Omega_{e}} \rho A \boldsymbol{\ddot{r}} \cdot \boldsymbol{\delta r} \:ds
    + \sum_{e=1}^{E} \int_{\Omega_{e}} \boldsymbol{f_{||}}    \cdot \boldsymbol{\delta r}^{\prime} \:ds 
    + \sum_{n=1}^{E-1}\:\Big[\langle\boldsymbol{f}\rangle \cdot \jump{\boldsymbol{\delta r}} \Big]\Big|_{s_{n}}
	+ \sum_{n=1}^{E-1}\:\Big[\jump{\boldsymbol{f}}\cdot \langle\boldsymbol{\delta r}\rangle \Big]\Big|_{s_{n}}\\
    &+ \sum_{e=1}^{E} \int_{\Omega_{e}} \boldsymbol{m_{\perp}} \cdot \boldsymbol{\delta \theta}^{\prime}\:ds 
    + \sum_{n=1}^{E-1}\:\Big[\langle\boldsymbol{m_{\perp}}\rangle \cdot \jump{\boldsymbol{\delta \theta}} \Big]\Big|_{s_{n}}
    + \sum_{n=1}^{E-1}\:\Big[\jump{\boldsymbol{m_{\perp}}}\cdot \langle\boldsymbol{\delta \theta}\rangle \Big]\Big|_{s_{n}}\\
    & = 
     \sum_{e=1}^{E} \int_{\Omega_{e}} \boldsymbol{\tilde{f}} \cdot \boldsymbol{\delta r} \:ds 
     + \left(\boldsymbol{\bar{f}} \cdot \boldsymbol{\delta r} \right)\Big|_{\partial_{N_f}\Omega_h} 
     + \sum_{e=1}^{E} \int_{\Omega_{e}} \boldsymbol{\tilde{m}_{\perp}} \cdot \boldsymbol{\delta \theta}\:ds
     + \left(\boldsymbol{\bar{m}_{\perp}} \cdot \boldsymbol{\delta \theta} \right)\Big|_{\partial_{N_m}\Omega_h},
\end{aligned}
\end{equation}

\noindent
Since jumps in forces $\boldsymbol{f}$ and bending moments $\boldsymbol{m_{\perp}}$ need not be
penalized to ensure the consistency of the numerical scheme, the terms involving their jumps in 
Equation~\eqref{eq:dg-weak-form} can be ignored, see also \cite{noels2006general,becker2011fracture}. 
However, the inter-element compatibility has to be enforced weakly to ensure the stability of the 
numerical scheme. Here, we do so through the interior penalty method, 
following~\cite{noels2006general,becker2011fracture}. Specifically, we add the following terms to 
the left-hand side of Equation \eqref{eq:dg-weak-form}:

$$
\sum_{n=1}^{E-1} \beta_{p}\Big[\bigg\langle\frac{EA}{h}\bigg\rangle\jump{\boldsymbol{\delta r}}\cdot\jump{\boldsymbol{r}}\Big]\Big|_{s_{n}} 
+ \sum_{n=1}^{E-1} \beta_{t}\Big[\bigg\langle\frac{EI}{h}\bigg\rangle\jump{\boldsymbol{\delta g_1}}\cdot\jump{\boldsymbol{g_1}}\Big]\Big|_{s_{n}},
$$
where we recall that $\boldsymbol{g_1} = \frac{\boldsymbol{r}^\prime}{||\boldsymbol{r}
^\prime||}$ (see Equation \eqref{eq:kirchhoff-constraint}), while $\beta_{p}>1$ and $\beta_{t}>1$ 
are position and tangent jump penalty parameters and $h$ is the element size.

We, therefore, obtain the following stabilized discontinuous Galerkin weak form:
\begin{equation}\label{eq:dg-weak-form-stab}
    \begin{aligned}
    \int_{\Omega_{h}} \rho A \boldsymbol{\ddot{r}} \cdot \boldsymbol{\delta r} \:ds
    &+ \int_{\Omega_{h}} EA\:\varepsilon \: \delta\varepsilon \:ds 
    + \sum_{n=1}^{E-1}\:\Big[\langle\boldsymbol{f}\rangle \cdot \jump{\boldsymbol{\delta r}} \Big]\Big|_{s_{n}}
    + \sum_{n=1}^{E-1} \beta_{p}\Big[\bigg\langle\frac{EA}{h}\bigg\rangle\jump{\boldsymbol{r}}\cdot\jump{\boldsymbol{\delta r}}\Big]\Big|_{s_{n}} \\
    &+ \int_{\Omega_{h}} EI\:\boldsymbol{\kappa} \cdot \boldsymbol{\delta \kappa} \:ds 
    + \sum_{n=1}^{E-1}\:\Big[\langle\boldsymbol{m_{\perp}}\rangle \cdot \jump{\boldsymbol{\delta \theta}} \Big]\Big|_{s_{n}}
    + \sum_{n=1}^{E-1} \beta_{t}\Big[\bigg\langle\frac{EI}{h}\bigg\rangle\jump{\boldsymbol{g_1}}\cdot \jump{\boldsymbol{\delta g_1}}\Big]\Big|_{s_{n}}\\
    = 
     \int_{\Omega_{h}} \boldsymbol{\tilde{f}} &\cdot \boldsymbol{\delta r} \:ds 
     + \left(\boldsymbol{\bar{f}} \cdot \boldsymbol{\delta r} \right)\Big|_{\partial_{N_f}\Omega_h}
     + \int_{\Omega_{h}} \boldsymbol{\tilde{m}_{\perp}} \cdot \boldsymbol{\delta \theta}\:ds
     + \left(\boldsymbol{\bar{m}_{\perp}} \cdot \boldsymbol{\delta \theta} \right)\Big|_{\partial_{N_m}\Omega_h},
\end{aligned}
\end{equation}
where we have also applied the constitutive relations \eqref{eq:constitutive}.

\subsection{The discontinuous Galerkin / cohesive zone model (DG/CZM) weak formulation}

As discussed in Section \ref{subsec:dg-formulation}, within the discontinuous Galerkin formulation, 
the position and tangent fields are allowed to exhibit discontinuities at the 
boundaries of the finite elements. However, prior to fracture, the solution's compatibility across 
element boundaries is maintained via the variationally-consistent interface forces and moments 
derived in that section.

Upon satisfaction of the fracture criterion~\eqref{eq18}, the interface axial forces and bending 
moments in Equation \eqref{eq:dg-weak-form-stab} are replaced with the cohesive axial forces 
and the cohesive bending moments of Equation \eqref{eq:cohesive-forces-moments}.
However, because we assume negligible shear strains and, consistently, we do not model the shear 
modes of fracture, we ensure that the component of the interface force term perpendicular to the unit 
normal of the cohesive boundary in Equation \eqref{eq:dg-weak-form-stab} remains active 
until complete interface failure~\cite{becker2011fracture}.
Specifically, this is achieved by introducing two binary parameters $\alpha_n$ and 
$\gamma_n$, which take value at each interface $n$. While we set both $\alpha_n = 1$ and 
$\gamma_n = 1$ before fracture initiation, we set $\alpha_n = 0$ after the fracture 
criterion~\eqref{eq18} is met, and $\gamma_n = 0$ upon complete decohesion (i.e. $\Delta \geq \Delta_{c}$).

The discontinuous Galerkin / cohesive zone model weak form reads, therefore: 
\begin{equation}\label{eq:dg-czm-weak-form}
    \begin{aligned}
    \int_{\Omega_{h}} \rho A \boldsymbol{\ddot{r}} \cdot \boldsymbol{\delta r} \:ds
    & + \int_{\Omega_{h}} EA\:\varepsilon \: \delta\varepsilon \:ds
    + \sum_{n=1}^{E-1}\:\alpha_n\Big[\boldsymbol{f_{DG,\:||}}\cdot\jump{\boldsymbol{\delta r}}\Big]\Big|_{s_{n}}
    + \sum_{n=1}^{E-1}\:\gamma_n\Big[\boldsymbol{f_{DG,\:\perp}}\cdot\jump{\boldsymbol{\delta r}}\Big]\Big|_{s_{n}} \\
    & + \sum_{n=1}^{E-1}\:\alpha_n\Bigg[\beta_{p}\bigg\langle\frac{EA}{h}\bigg\rangle\:\boldsymbol{c_{DG,\:||}}\cdot \jump{\boldsymbol{\delta r}}\Bigg]\Bigg|_{s_{n}} 
    + \sum_{n=1}^{E-1}\:\gamma_n\Bigg[\beta_{p}\bigg\langle\frac{EA}{h}\bigg\rangle\:\boldsymbol{c_{DG,\:\perp}}\cdot\jump{\boldsymbol{\delta r}}\Bigg]\Bigg|_{s_{n}} \\
    & + \int_{\Omega_{h}} EI\:\boldsymbol{\kappa} \cdot \boldsymbol{\delta \kappa} \:ds 
    + \sum_{n=1}^{E-1}\:\alpha_n\Big[\langle\boldsymbol{m_{\perp}}\rangle \cdot \jump{\boldsymbol{\delta \theta}} \Big]\Big|_{s_{n}}
    + \sum_{n=1}^{E-1}\:\alpha_n\Bigg[\beta_{t}\bigg\langle\frac{EI}{h}\bigg\rangle\jump{\boldsymbol{g_1}}\cdot\jump{\boldsymbol{\delta g_1}}\Bigg]\Bigg|_{s_{n}} \\
    & + \sum_{n=1}^{E-1}\:(1-\alpha_n)\Big[\boldsymbol{f_{coh,\:||}}\cdot\jump{\boldsymbol{\delta r}}\Big]\Big|_{s_{n}}
	+ \sum_{n=1}^{E-1}\:(1-\alpha_n)\Big[\boldsymbol{m_{coh,\:\perp}}\cdot\jump{\boldsymbol{\delta g_1}}\Big]\Big|_{s_{n}} \\
    = \int_{\Omega_{h}} \boldsymbol{\tilde{f}} &\cdot \boldsymbol{\delta r} \:ds 
    + \left(\boldsymbol{\bar{f}} \cdot \boldsymbol{\delta r} \right)\Big|_{\partial_{N_f}\Omega_h}
    + \int_{\Omega_{h}} \boldsymbol{\tilde{m}_{\perp}} \cdot \boldsymbol{\delta \theta}\:ds
    + \left(\boldsymbol{\bar{m}_{\perp}} \cdot \boldsymbol{\delta \theta} \right)\Big|_{\partial_{N_m}\Omega_h}
    \end{aligned}
\end{equation}

where we have set:

\begin{equation*}
    \begin{aligned}
    \boldsymbol{f_{DG,\:||}} &=
	\left(\langle\boldsymbol{f}\rangle\cdot\boldsymbol{\hat{n}_{coh}}\,\right)\,\boldsymbol{\hat{n}_{coh}},\\
    \boldsymbol{f_{DG,\:\perp}} &= \left(\boldsymbol{I} -
		\boldsymbol{\hat{n}_{coh}}\otimes\boldsymbol{\hat{n}_{coh}}\right)\langle\boldsymbol{f}\rangle,\\
\boldsymbol{c_{DG,\:||}} &=
	\left(\jump{\boldsymbol{r}}\cdot\boldsymbol{\hat{n}_{coh}}\,\right)\,\boldsymbol{\hat{n}_{coh}},\\
\boldsymbol{c_{DG,\:\perp}} &= \left(\boldsymbol{I} -
		\boldsymbol{\hat{n}_{coh}}\otimes\boldsymbol{\hat{n}_{coh}}\right)\jump{\boldsymbol{r}}.
    \end{aligned}
\end{equation*}

Finally, in the event of recontact of the cracked surfaces after fracture initiation 
(i.e. when $\Delta_{||} < 0$ and $\Delta_{max} > 0$), we reactivate the variationally-consistent 
axial forces and the position stabilization term, so as to allow propagation of compressive stress 
waves across cracked interfaces.

\pagestyle{plain}

\subsection{Discretization in space and time} \label{sec:space-time}

We discretize $\boldsymbol{r}$ in space with third-order Hermite polynomials, which are shown to 
yield convergence order of four in continuous Galerkin settings, see \cite{meier2019geometrically}.
The preference for Hermite polynomials over Lagrange polynomials stems from the advantage that 
the arc-length derivative $\boldsymbol{r^{\prime}}$ at each node 
is directly available as a primary degree of freedom associated with that particular node,
which simplifies the computation of the terms at the element interfaces in Equation 
(28). We therefore discretize $\boldsymbol{r}$ as follows:
\begin{equation}\label{eq26}
	\boldsymbol{r}(\xi) \approx \sum_{b=1}^{2}N_{p}^{b}(\xi)\boldsymbol{p}^{b} + \frac{L}{2}\sum_{b=1}^{2}N_{t}^{b}(\xi)\boldsymbol{t}^{b} := \sum_{a=1}^{4}N_{a}(\xi)\boldsymbol{x_{a}},
\end{equation}
where $\boldsymbol{p}^{b}$ and $\boldsymbol{t}^{b}$ are the position and tangent degrees of freedom 
at the element nodes $b = 1,2$, $\xi \in [-1,1]$ is the parametric coordinate in the reference 
element, which is mapped to the arc-length coordinate $s \in [0,L]$ as $s = \frac{L}{2}(1+\xi)$, and 
$N_{p}^{b}$ and $N_{t}^{b}$ are the following shape functions:

\begin{equation*}
    \begin{aligned}
	N_{p}^{1}(\xi) &= \frac{1}{4}(2+\xi)(1-\xi)^2\:; \:\: N_{p}^{2}(\xi) = \frac{1}{4}(2-\xi)(1+\xi)^2, \\
	N_{t}^{1}(\xi) &= \frac{1}{4}(1+\xi)(1-\xi)^2\:; \:\: N_{t}^{2}(\xi) = -\frac{1}{4}(1-\xi)(1+\xi)^2.
    \end{aligned}
\end{equation*}

This space discretization results in the following semi-discrete system of equations:
\begin{equation}\label{eq30}
	\boldsymbol{M_{ab}}\boldsymbol{\ddot{x}_{b}} + \boldsymbol{f^{int}_{a}} + \boldsymbol{f^{jump}_{a^{\pm}}} = \boldsymbol{f^{ext}_{a}},
\end{equation}
where the inertia $\boldsymbol{M_{ab}}\boldsymbol{\ddot{x}_{b}}$, internal (bulk) $\boldsymbol{f^{int}_{a}}$, 
internal (interface) $\boldsymbol{f^{jump}_{a^{\pm}}}$, and external $\boldsymbol{f^{ext}_{a}}$ 
forces are reported in \ref{sec:appendixA}.

\noindent
Given the strong nonlinearities involved with fracture, we opted for an explicit time discretization
for simplicity.
Specifically, we discretize Equation~\eqref{eq30} in time using the second-order explicit Newmark scheme.
We perform a special mass lumping~\cite{hughes2012finite} to avoid the cost of solving a linear 
system for computing the accelerations. As explicit time stepping schemes are conditionally stable, 
we compute the stable time step $\Delta t_{c}$ from the Courant-Friedrichs-Lewy 
condition:
\begin{equation}\label{eq34}
	\Delta t_{c} = \frac{2}{\omega_{max}},
\end{equation}
where $\omega_{max} = \max_{i=1}^{N}(||\lambda_{i}||)$ is the maximum natural frequency of the 
system. 

\noindent
In our simulations, we compute the stable time step $\Delta t_{c}$ only once at the beginning of the 
calculation and we set $\Delta t = \Delta t_{c}$ for the entire calculation. 
Specifically, we calculate $\Delta t_{c}$ by solving the following linearized eigenvalue problem:
\begin{equation}\label{eq35}
	\left(\boldsymbol{K_{ab}^{int}} + \boldsymbol{K_{ab^{\pm}}^{jump,DG}}-\lambda^2\:\boldsymbol{M_{ab}^{lump}}\right)\boldsymbol{\Phi} = \boldsymbol{0},
\end{equation}
where $\lambda$ and $\boldsymbol{\Phi}$ are the eigenvalue and eigenvector pair, 
$\boldsymbol{M_{ab}^{lump}}$ is the lumped mass matrix, and the stiffness matrices $\boldsymbol{K_{ab}^{int}}$ 
and $\boldsymbol{K_{ab^{\pm}}^{jump,DG}}$ are reported in \ref{sec:appendixB}.
It should be noted that the eigenvalues obtained by solving Equation~\eqref{eq35} 
may be complex since matrix $\boldsymbol{K_{ab^{\pm}}^{jump,DG}}$ is not symmetric.

\pagestyle{plain}

\section{Convergence analysis of the discontinuous Galerkin discretization}

This section presents a convergence analysis of the discontinuous Galerkin space discretization 
in the geometrically nonlinear regime.

\subsection{Bending of a cantilever beam}

In this example, we analyze the convergence of the discontinuous Galerkin finite element discretization 
on a benchmark from Simo and Vu-Quoc~\cite{simo1986three}. We consider a cantilever beam of length $L$, 
radius $R$ and Young's modulus $E$ with a concentrated end moment of magnitude $M = \pi^2ER^4$/$L$, causing the 
beam to wind around itself twice forming a double circle. Table~\ref{tab:cantilever-beam-bending} 
summarizes the physical properties used in this problem.

{\def\arraystretch{1.25}\tabcolsep=15pt
\begin{table}[h!]
	\begin{center}
		\begin{tabular}{ c c }
			\hline
			Property                                	& 
			Value\\
			\hline
			Length ($L$)                                      & 
			$1 \:m$\\
			Radius ($R$)                                      & 
			$0.01 \:m$\\
			Young's modulus ($E$)                             			& 
			$200.0\:GPa$\\
			\hline
		\end{tabular}
        \caption{\textit{Physical properties used in the bending of a cantilever beam benchmark.}} \label{tab:cantilever-beam-bending}
	\end{center}
\end{table}
}

We consider uniform meshes with element sizes starting at 0.125 $m$ up to six 
levels of refinement, where each mesh refinement reduces the element size by a factor of 2. We also 
vary the penalty parameters ($\beta_{p,n}$ and $\beta_{t,n}$) amongst three values, namely 10, 100 and 1000, to 
investigate their influence on the convergence order of the framework. We solve this problem with our 
discontinuous Galerkin discretization in a quasi-static setting using a Newton-Raphson solver (linearization 
provided in \ref{sec:appendixB}) by applying the concentrated end moment in 50 load steps for each case. 
Following~\cite{meier2014objective}, we calculate the error in the deformed position of the beam centerline 
with the following relative $L^2$ norm:
\begin{equation}
	||e||^2_{rel} = \frac{1}{u_{max}}\sqrt{\frac{1}{L}\int_{0}^{L}||\boldsymbol{r}_{h}-\boldsymbol{r}_{ref}||^2\:ds}
\end{equation}

\noindent
where $\boldsymbol{r}_{h}$ and $\boldsymbol{r}_{ref}$ are the numerical and reference (analytical) beam centerline 
positions respectively and $u_{max}$ is the maximum displacement obtained in a particular case.

\begin{figure}[h!]
	\begin{center}
		\includegraphics[width=0.75\textwidth]{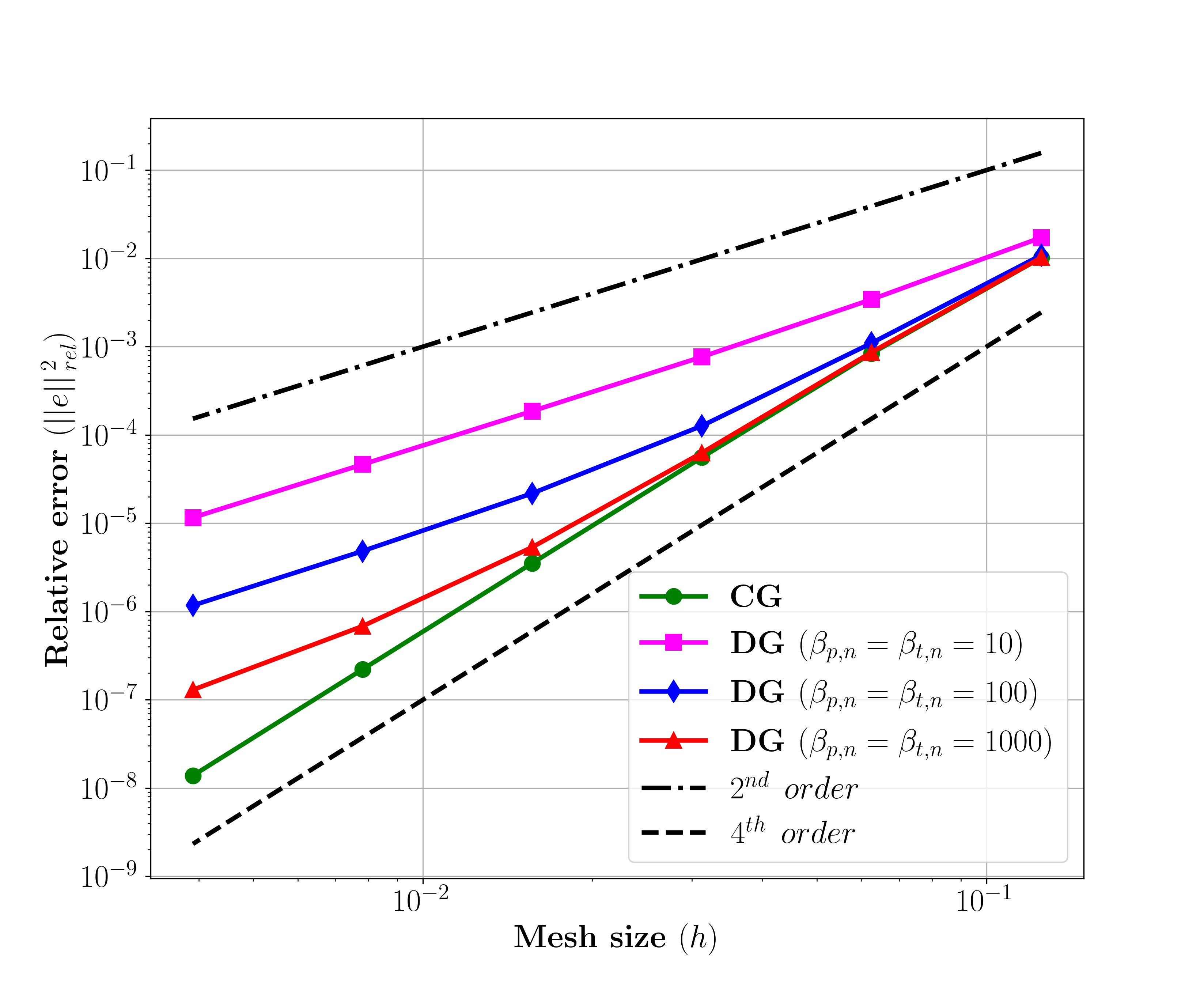}
	\end{center}
	\caption{\textit{
        Convergence analysis of the discontinuous Galerkin (DG) finite element discretization for 
        the bending of a cantilever beam benchmark. The plot shows the relative $L^2$ error in the 
        deformed beam centerline positions as a function of the mesh size for different penalty 
        parameter values ($\beta_{p,n}$ and $\beta_{t,n}$).
        Reference lines representing the convergence orders two and four are shown with dotted lines. 
        The convergence order of the DG discretization approaches four, which is the expected value 
        for a continuous Galerkin (CG) discretization, for higher values of the penalty parameters. 
        Convergence deteriorates for smaller values of the penalty parameters.}}
	\label{fig:convergenceplots}
\end{figure}

The results of the convergence analysis are shown in Figure~\ref{fig:convergenceplots}. We observe 
that the convergence order of the discontinuous Galerkin finite element discretization approaches the expected 
value of four, as obtained in the continuous Galerkin setting~\cite{meier2014objective}, for 
increasing values of the penalty parameters, while it deteriorates for smaller values of $\beta_{p,n}$ and $\beta_{t,n}$.
Our results are consistent with the findings by Brezzi et al.~\cite{brezzi2000discontinuous} that 
suggest the need for large penalty parameters to achieve the expected order of convergence in penalty 
based DG methods, especially for higher order polynomials. However,
in implicit analyses, large penalty parameters lead to a large condition number of the stiffness 
matrix~\cite{brezzi2000discontinuous}, resulting in an increased number of iterations for convergence of iterative linear 
solvers and in a larger simulation runtime. 
Conversely, in explicit analyses, large penalty parameters result 
in smaller stable time steps~\cite{noels2008explicit}, thus also increasing the simulation runtime. 
Therefore,
$\beta_{p,n}$ and $\beta_{t,n}$ should be chosen considering the usual 
tradeoff between accuracy and computational time. Nevertheless, a relative 
error of less than 1\% and a consistent reduction in the error with h-refinement is observed across all 
examined mesh sizes and penalty parameter values.

\section{Results} \label{sec:results}

This section presents verification and validation of our computational framework.

\subsection{Framework verification: Buckling of a slender column}

We first verify our computational framework in the absence of fracture for the case of a
column buckling under quasi-static axial compression.
Specifically, we consider a slender column of length $L$, radius $R$ and Young's modulus $E$
and we verify that the first buckling mode occurs at Euler's critical axial force:
\begin{equation}\label{beam-buckling-load}
	f_{cr} = \frac{\pi^3}{4} \frac{ER^4}{L^2}.
\end{equation}

\begin{figure}[h!]
	\begin{center}
		\includegraphics[width=0.135\textwidth]{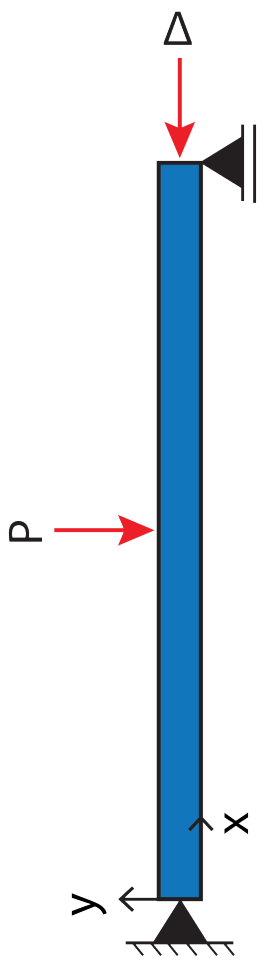}
	\end{center}
	\caption{\textit{Schematic of the geometry and boundary conditions for the slender column
    buckling benchmark.
    The column is supported with a pin at the bottom end, a roller at the top end, and is loaded
    with an imposed axial displacement $\Delta$. To facilitate the occurrence of the buckling
    bifurcation, a small, transverse perturbation force $P$ is applied at the center of the column
    to break the problem's symmetry.}} \label{fig4}
\end{figure}

The problem geometry and boundary conditions considered are illustrated in Figure~\ref{fig4}.
We apply a pin support at the bottom end of the column ($x = y = 0$), and a roller support
allowing displacement in the $x$-direction at the top end ($x = L\:,\:y = 0$), where we also impose
a displacement in the negative $x$ direction, ramping up quasi-statically to a maximum $\Delta$
through 1000 load steps.
In addition, we apply a constant perturbation force $P$ at the center of the column
($x = L/2\:,\:y = 0$) as a means to break the symmetry of the problem.
Table \ref{tab:buckling} summarizes the physical properties and numerical parameters used.

{\def\arraystretch{1.25}\tabcolsep=15pt
\begin{table}[h!]
	\begin{center}
		\begin{tabular}{ c c }
			\hline
			Property / Parameter                                 & Value\\
			\hline
			Length ($L$)                                      & $10 \:m$\\
			Radius ($R$)                                      & $0.1 \:m$\\
			Young's modulus ($E$)                             & $200.0\:GPa$\\
			Final applied displacement ($\Delta$)             & $5.0\:mm$\\
			Applied perturbation force ($P$)                & $1.0\:N$\\
			Mesh size ($h$)                                      & $1.0 \:m$\\
			Penalty parameters ($\beta_{p,n}$ and $\beta_{t,n}$) & 10\\
			\hline
		\end{tabular}
        \caption{\textit{Physical properties and numerical parameters used in the slender column
buckling benchmark.}} \label{tab:buckling}
	\end{center}
\end{table}
}

\begin{figure}[h!]
	\begin{center}
		\includegraphics[width=0.60\textwidth]{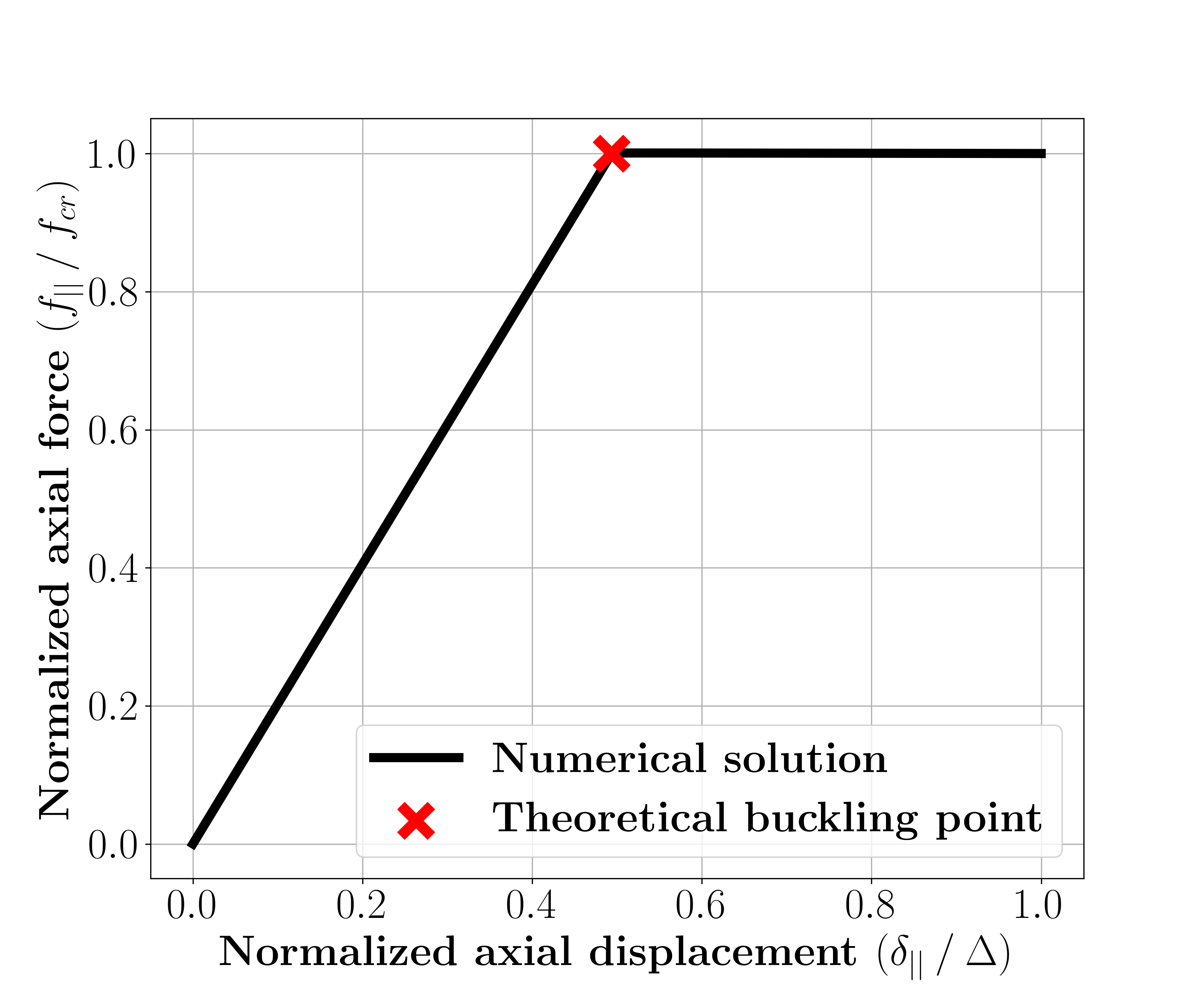}
	\end{center}
	\caption{\textit{Verification of our computational framework in the case of buckling of a slender
    column. The plot shows the force-displacement response at the top end of the column.
    The axial force $f_{||}$ is normalized with respect to the theoretical critical axial force $f_{cr}$, while
    the axial displacement $\delta_{||}$ is normalized with respect to the final applied displacement $\Delta$.
    The theoretical buckling point is marked with a cross.}}
	\label{fig5}
\end{figure}

We apply our DG/CZM computational framework to solve the problem in a quasi-static setting
using a Newton-Raphson solver with the linearization provided in \ref{sec:appendixB}.
Figure~\ref{fig5} shows that the load-displacement response of the column is in agreement with the
theoretical predictions.

\FloatBarrier
\subsection{Framework verification: Fracture of a slender bar under tension}

\begin{figure}[h!]
	\begin{center}
		\includegraphics[width=0.60\textwidth]{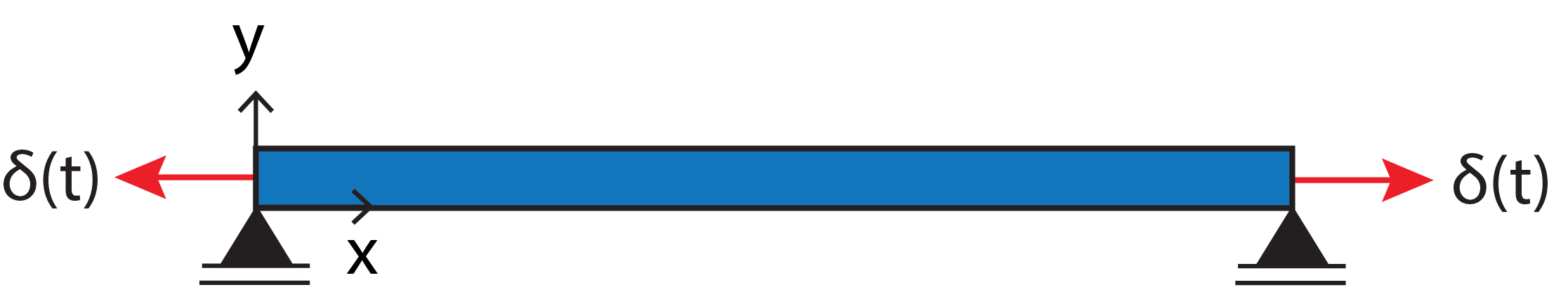}
	\end{center}
	\caption{\textit{Schematic of the geometry and boundary conditions for the bar spall benchmark.
    The axial displacement signal $\delta(t)$ in Equation \eqref{eq37} is applied at the two ends
    of the bar, while two roller supports restrain the transverse displacement.}}
	\label{fig6}
\end{figure}

We verify the fracture modeling capability of our computational framework by simulating the spall
of a bar, i.e. the development of a crack as a result of the interaction of two tensile
stress waves.
We consider a slender bar subjected to tensile axial loading, while transverse displacement is
restrained through roller supports at the bar ends, as illustrated in Figure~\ref{fig6}.
We apply the following axial displacement signal:
\begin{equation}\label{eq37}
	\delta(t) = \frac{\sigma_f}{2}\frac{t}{\rho c_{l}},
\end{equation}
where $c_{l} = \sqrt{E/\rho}$ is the bar longitudinal wave speed and $\sigma_f$ is a stress loading
factor.
Table~\ref{tab1} summarizes the physical properties and the numerical parameters used in this
computational experiment.

{\def\arraystretch{1.25}\tabcolsep=15pt
	\begin{table}[htb!]
		\begin{center}
			\begin{tabular}{ c c }
				\hline
				Property / Parameter                                 & Value \\
				\hline
                Length ($L$)                                    & $0.1 \:m$\\
                Radius ($R$)                                    & $1.0 \:mm$\\
                Mass density ($\rho$)                                & $3690\:kg\ m^{-3}$ \\
				Young's modulus (E)                                  & $260.0\:GPa$ \\
				Critical cohesive strength ($\sigma_c$)              &  $400.0\:MPa$ \\
				Fracture energy ($G_c$)                              & $100.0\:N\ m^{-1}$ \\
				Mode-mixity parameter ($\alpha$)                     & $1$ \\
				Simulation time ($T$)                                & $10\:\mu s$ \\
				Mesh size ($h$)                                      & $0.1 \:mm$ \\
				Time step ($\Delta t$)                               & $0.01\:ns$ \\
				Penalty parameters ($\beta_{p,n}$ and $\beta_{t,n}$) & $10$ \\
				\hline
			\end{tabular}
			\caption{\textit{Physical properties and numerical parameters used in the bar spall
            benchmark.}}
			\label{tab1}
		\end{center}
	\end{table}
}

\begin{figure}[h!]
	\begin{center}
		\includegraphics[width=0.75\textwidth]{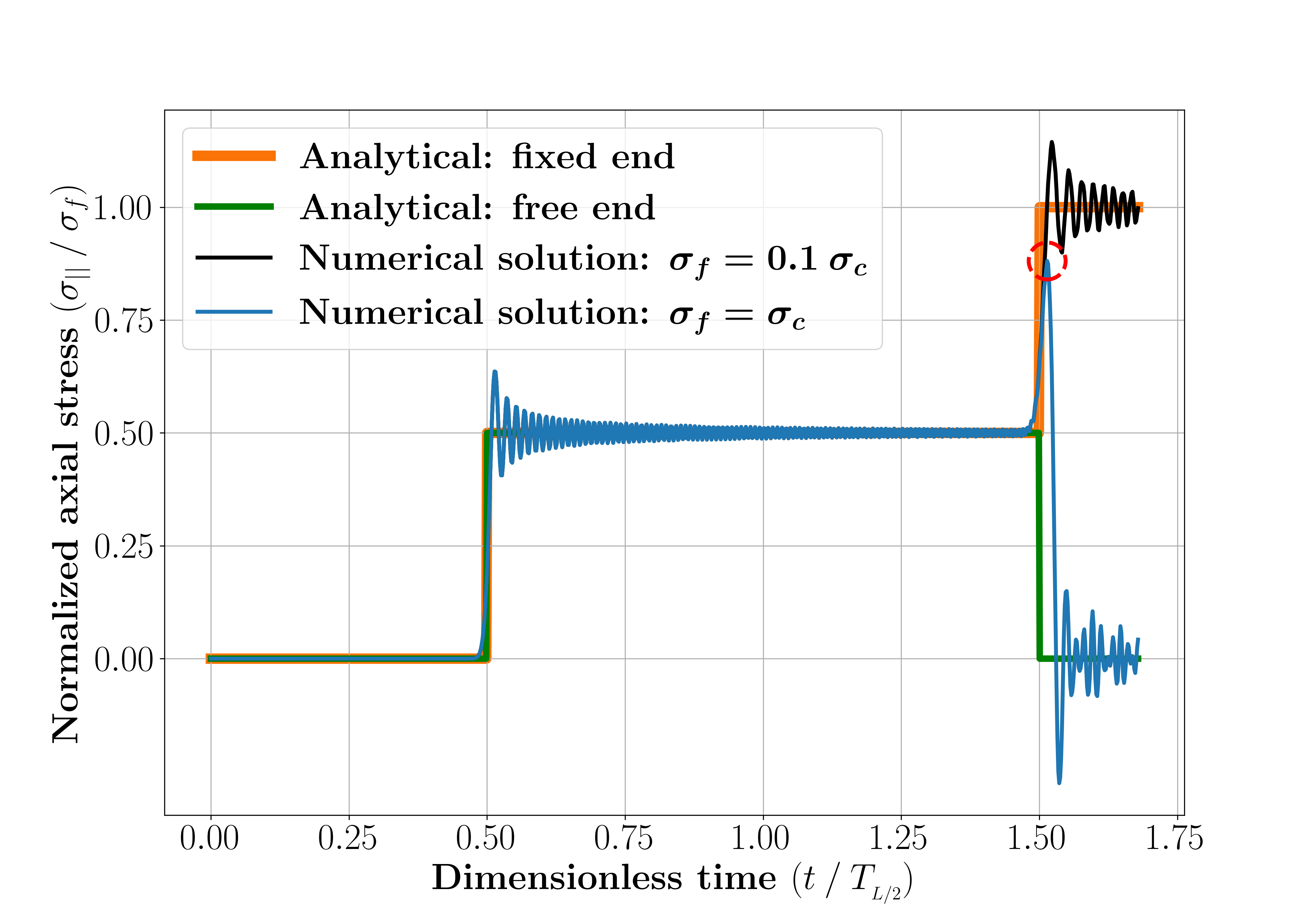}
	\end{center}
	\caption{\textit{Verification of our computational framework applied to the fracture of a slender bar
    under tensile load.
    The plot shows a comparison of the simulated and theoretical axial stress evolution in
    time, gauged at a point $L/4$ from the bar end, for two different applied displacement
    signals (Equation \eqref{eq37}), one with $\sigma_f < \sigma_c$ (no fracture) and the other
    with $\sigma_f = \sigma_c$ (fracture).
    The theoretical predictions correspond to the solution of the 1D wave equation on a bar of half
    length with the same applied displacement signal on one end and the other end fixed (no
	fracture) or free (fracture).
    The axial stress $\sigma_{||} = E \varepsilon$ is normalized with respect to the applied stress
    $\sigma_f$, while the time is normalized with respect to $T_{L/2} = L/(2c_{l})$,
    which is the time when the longitudinal stress waves meet at the center of the bar.
    When $\sigma_f < \sigma_c$, our numerical results closely match the theoretical response of an uncracked bar. Conversely, at $\sigma_f = \sigma_c$, our results accurately capture the fracture event at 
	$t = T_{L/2}$ due to the interaction of two tensile waves, the development of the release wave (see red 
	circle) followed by the vanishing of the axial force due to fully developed fracture at the center of the beam.
    }}
	\label{fig7}
\end{figure}

\begin{figure}[h!]
	\centering
	\begin{subfigure}{0.35\textwidth}
		\centering
		\includegraphics[width=0.65\textwidth]{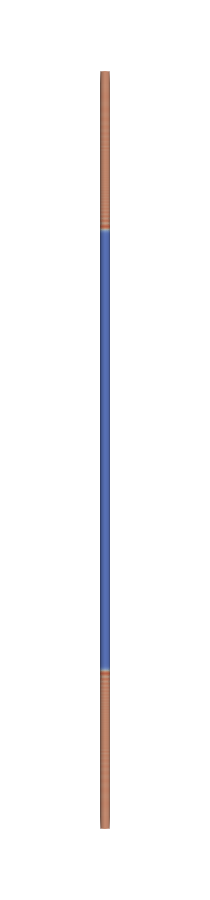}
		\vspace{-1.0cm}
		\caption{}
	\end{subfigure}
	\hspace{-3.0cm}
	\begin{subfigure}{0.35\textwidth}
		\centering
		\includegraphics[width=0.65\textwidth]{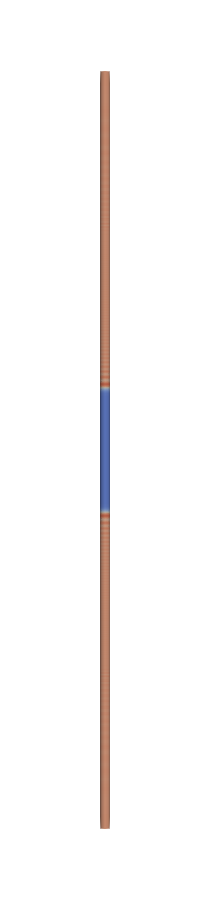}
		\vspace{-1.0cm}
		\caption{}
	\end{subfigure}
	\hspace{-3.0cm}
	\begin{subfigure}{0.35\textwidth}
		\centering
		\includegraphics[width=0.65\textwidth]{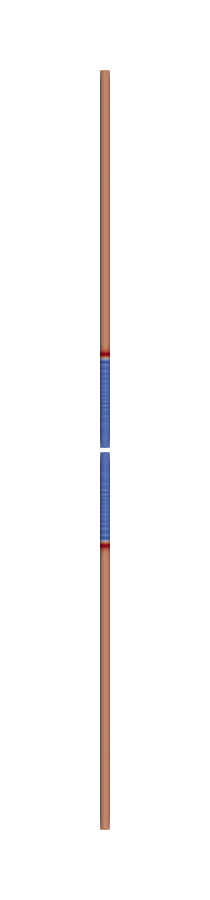}
		\vspace{-1.0cm}
		\caption{}
	\end{subfigure}
	\hspace{-3.0cm}
	\begin{subfigure}{0.35\textwidth}
		\centering
		\includegraphics[width=0.65\textwidth]{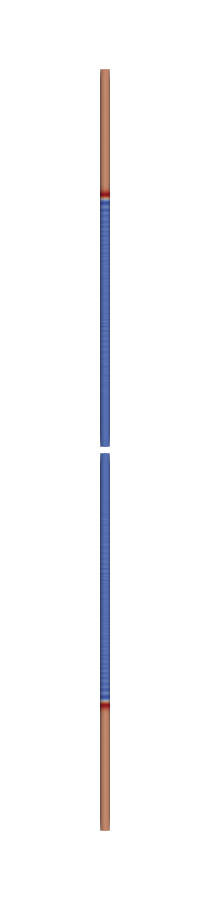}
		\vspace{-1.0cm}
		\caption{}
	\end{subfigure}
	\hspace{-2.5cm}
	\begin{subfigure}{0.25\textwidth}
		\centering
		\includegraphics[width=0.75\textwidth]{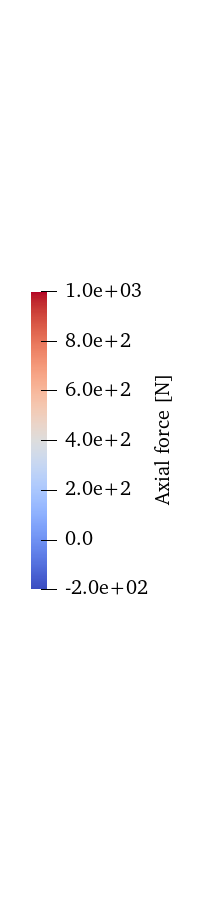}
	\end{subfigure}
	\caption{\textit{Snapshots of pre-fracture and post-fracture evolution of the bar deformation
        and axial force at times:
		(a) $0.42 \ T_{L/2}$, (b) $0.84 \ T_{L/2}$, (c) $1.26  \ T_{L/2}$, and (d) $1.68 \ T_{L/2}$.
        The bar is shown in its deformed configuration
        (displacements are scaled by a factor of 10 for better visualization) and the contours
        show the axial forces in the bar.
        We observe the propagation of the two tensile step waves towards the center of the bar
        (pictures (a) and (b)), the creation of a fracture surface as a result of their interaction
        at the center, as well as the reflection of the stress waves at the newly created free
        surface (pictures (c) and (d)).
        }}
	\label{fig8}
\end{figure}

We apply our DG/CZM computational framework to model the dynamic response of the bar with two values
of $\sigma_f$, namely $\sigma_f = 0.1 \ \sigma_c$ and $\sigma_f = \sigma_c$.
Figure~\ref{fig7} presents the evolution of the axial stress response over time obtained in our
simulations, against the corresponding analytical solution to the 1D wave equation.
It can be observed that the applied displacement signal generates two tensile stress waves with step
waveform and intensity $\sigma_f/2$ propagating inwards from the beam ends.
The two tensile waves meet at the center of the bar, building up a tensile stress wave of intensity
$\sigma_f$.
When $\sigma_f$ is less than the critical fracture strength $\sigma_c$, our results show that the
bar remains undamaged, as expected.
When $\sigma_f$ is equal to the critical fracture strength $\sigma_c$, our results capture the
fracture response, including the development of a release wave and the subsequent vanishing of the
axial stress due to the creation of a free surface.
Figure~\ref{fig8} presents snapshots of the simulated bar response showing the bar deformation the
propagation of the stress waves in the bar before and after fracture.

\FloatBarrier
\subsection{Fracture of a bar under transverse load}

Through this example, we demonstrate the significance of incorporating inter-element jumps in the
tangent (rotational) degrees of freedom and of modeling the relaxation of the bending moments as a function of
these jumps, in the spirit of the well-established traction-separation laws of fracture mechanics.
This is in contrast to the computational framework for fracture in geometrically exact beams
proposed by Tojaga et al.~\cite{tojaga2023geometrically}, where only the displacement degrees
of freedom are enriched with discontinuous modes, but not the rotational ones.

\begin{figure}[h!]
	\begin{center}
		\includegraphics[width=0.425\textwidth]{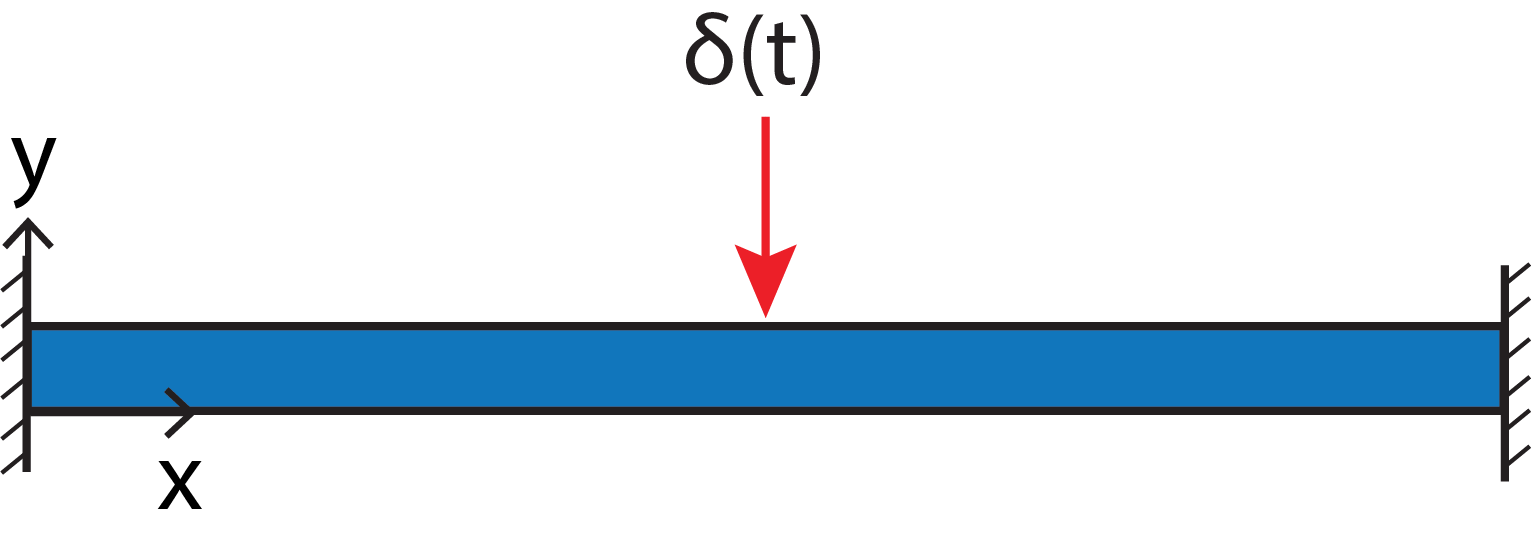}
	\end{center}
	\caption{\textit{Schematic of the geometry and boundary conditions for the problem of a bar
        under transverse load. The beam is cantilevered at both ends and loaded at the center with
        a time-dependent transverse displacement $\delta(t)$, see Equation
        \eqref{eq:transverse-loading}.}}
	\label{fig9}
\end{figure}

We consider a slender bar cantilevered at both ends and loaded at the center, as illustrated in
Figure \ref{fig9}.
Specifically, we apply a time-dependent transverse displacement $\delta(t)$:
\begin{equation}\label{eq:transverse-loading}
\delta(t) = \tilde{v} t
\end{equation}
where $\tilde{v}$ is a sufficiently low loading rate to achieve quasi-static conditions.
Table~\ref{tab2} summarizes the physical properties and the numerical parameters used in this
computational experiment.

{\def\arraystretch{1.25}\tabcolsep=15pt
	\begin{table}[htb!]
		\begin{center}
			\begin{tabular}{ c c }
				\hline
				Property / Parameter                            & Value \\
				\hline
                Length ($L$)                                    & $0.1 \:m$\\
                Radius ($R$)                                    & $1.0 \:mm$\\
                Mass density ($\rho$)                           & $3690\:kg\ m^{-3}$ \\
				Young's modulus (E)                             & $260.0\:GPa$ \\
				Critical cohesive strength ($\sigma_c$)         &  $400.0\:MPa$ \\
				Fracture energy ($G_c$)                         & $100.0\:N\ m^{-1}$ \\
				Mode-mixity parameter ($\alpha$)                & $1$ \\
				Load rate ($\tilde{v}$)                         & $0.01\:m\:s^{-1}$ \\
				Simulation time ($T$)                           & $0.2\:s$ \\
				Mesh size ($h$)                                 & $0.1 \:mm$ \\
				Time step ($\Delta t$)                          & $0.1\:\mu s$ \\
				Penalty parameters ($\beta_{p,n}$ and $\beta_{t,n}$) & $10$ \\
				\hline
			\end{tabular}
			\caption{\textit{Physical properties and numerical parameters used in the bar under
            transverse load problem.}}
			\label{tab2}
		\end{center}
	\end{table}
}

\begin{figure}[h!]
	\begin{center}
		\includegraphics[width=0.55\textwidth]{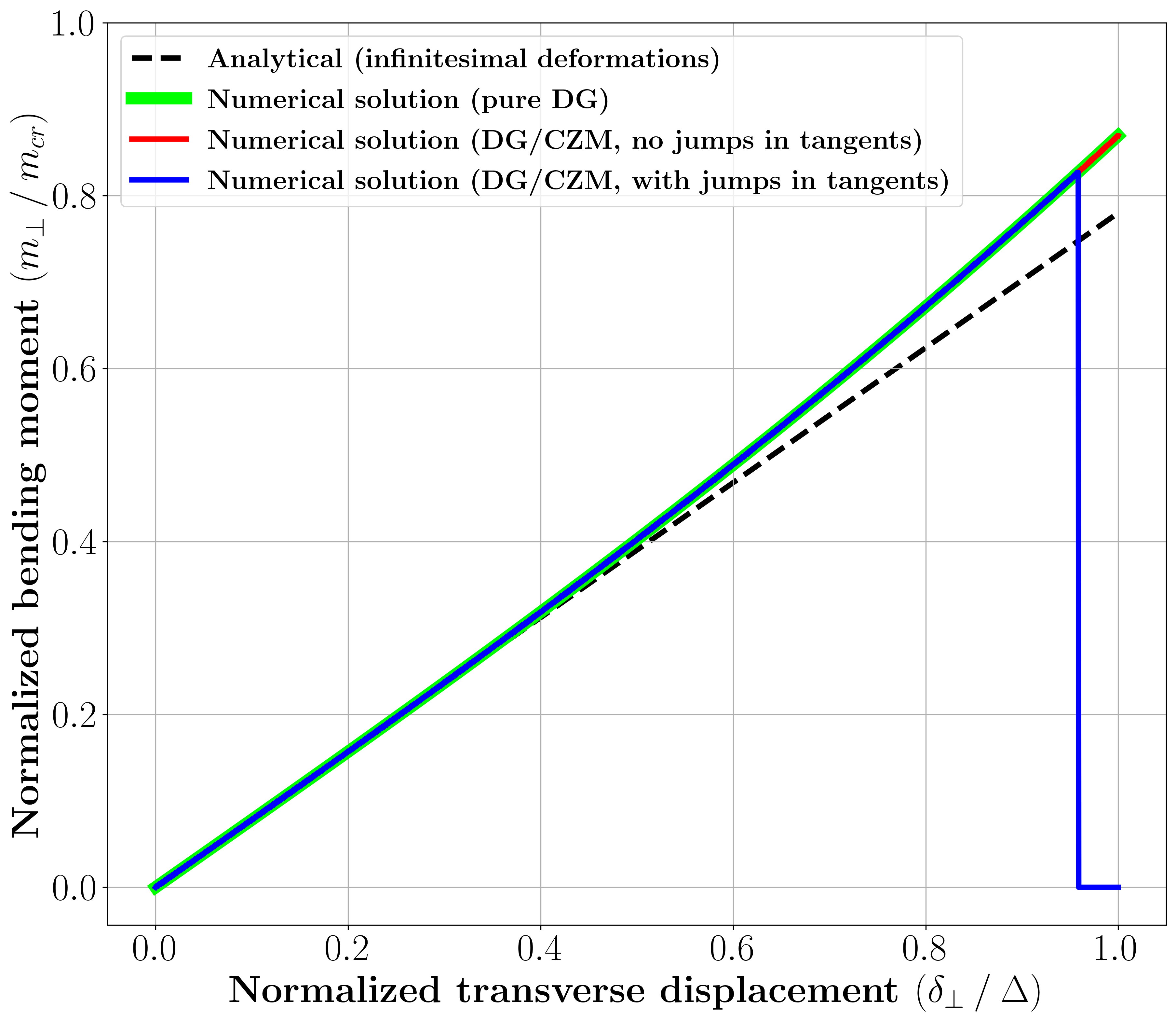}
	\end{center}
	\caption{\textit{The load-displacement response at the center of a bar fracturing under
    transverse load, in terms of bending moment versus applied transverse displacement.
    Simulation results are plotted on top of the analytical predictions of Euler-Bernoulli beam
    theory (dashed line) to highlight the departure from the geometrically linear regime.
    The bending moment is normalized with respect to the critical bending moment
    $m_{cr} = AR \sigma_c$, where $A$ is the area of the beam cross section, while the transverse
    displacement of the beam is normalized
    with respect to the maximum applied transverse displacement $\Delta = \tilde{v}T$.
	The plot compares simulation results from our DG/CZM framework with (blue line) and without 
	(red line) discontinuities in tangent degrees of freedom. In the former case, bar failure under bending 
	is accurately captured, resulting in a drop in bending moment, while in the latter case, fracture behavior 
	is absent, and the bending moment follows a response predicted by a simulation without an embedded 
	fracture mechanics model (green line).}}
	\label{fig11}
\end{figure}

We solve this problem computationally with two different approaches: 1) our DG/CZM computational
framework as described in Section \ref{sec:dg-formulation}, and 2) a variant of such framework that
does not model the bending moment decay with the increasing tangents jumps (i.e. removing the
bending moment term in Equation~\eqref{eq19}).
Figure~\ref{fig11} compares the load-displacement responses of the center of the bar obtained with these two approaches.
The figure shows the bending moments build-up with the increasing applied displacement, including
the transition from the geometrically linear to the geometrically nonlinear regime.
We observe that our DG/CZM approach is able to capture the fracture behavior arising from a further
increase in applied displacement, along with the vanishing bending moment in the post-fracture
behavior.
This stands in stark contrast to the second approach, which predicts a load-displacement response
identical to that of a simulation without an embedded fracture mechanics model (pure DG).
This demonstrates the essential role of explicitly modeling jumps in the tangent degrees of freedom
to capture the bending fracture mode.

We also observe that, under these conditions, fracture occurs in mixed bending and tensile modes,
as complete fracture is achieved for an internal moment $m_{\perp} \approx 0.8\ m_{cr} < m_{cr}$,
where $m_{cr} = AR \sigma_c$, where $A$ is the area of the beam cross section, see
Figure~\ref{fig11}.

\begin{figure}[h!]
	\centering
	\begin{subfigure}{0.85\textwidth}
		\centering
		\includegraphics[width=0.85\textwidth]{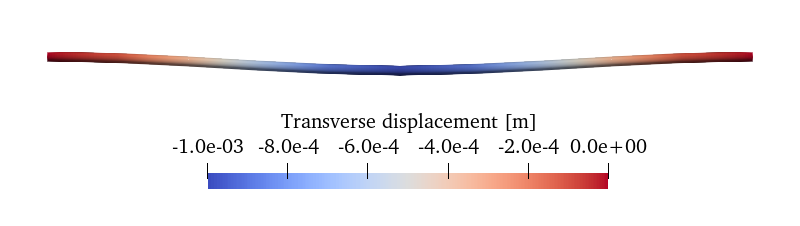}
	\end{subfigure}
	\begin{subfigure}{0.85\textwidth}
		\centering
		\includegraphics[width=0.85\textwidth]{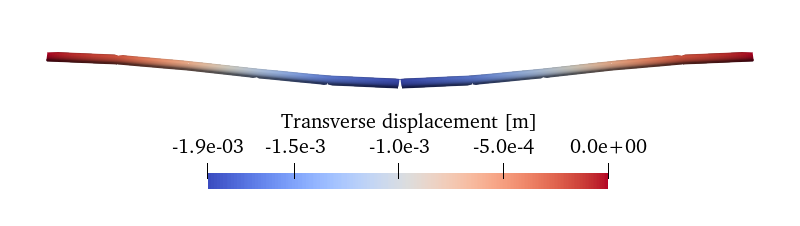}
	\end{subfigure}
	\caption{\textit{Snapshots of the simulated bar response before (top picture,
        $\delta(t) = 1.0\ mm$) and after (bottom picture, $\delta(t) = 1.92\ mm$) fracture.
        The bar is shown in its deformed configuration (displacements are scaled by a factor of 2
        for better visualization) and the contours show the transverse displacement.}}
	\label{fig10}
\end{figure}

Figure~\ref{fig10} presents snapshots of the simulated bar response showing the bar configuration
before and after fracture.

Note that in this physical scenario, the occurrence of fracture 
within the beam is anticipated both at its supports and at the center.
However, our computational framework does not describe fracture at the supports as it limits the 
possible fracture initiation sites to the inter-element boundaries.
While our computational framework can be adapted to model fracture at the beam's ends as well,
we did not pursue that route, as the purpose of this example was purely to illustrate 
the significance of incorporating inter-element jumps in tangent (rotational) degrees of freedom 
when modeling the bending mode of fracture.

\FloatBarrier
\subsection{Framework validation: Fracture of a bar bent and suddenly released}

In this section, we validate our computational framework against experiments by Audoly and
Neukirch~\cite{audoly2005fragmentation} on the fragmentation of dry spaghetti.
Specifically, that work presents experiments on dry spaghetti rods initially bent into an arc of
circle and suddenly released at one end, while the other end remains clamped.
The experiments show that this sudden relaxation of curvature results in a burst of flexural waves,
which locally increase the curvature of the rod, ultimately leading to fracture.

We model these experiments with a two-stage solution strategy.
As in experiments~\cite{audoly2005fragmentation}, the spaghetti rod remains clamped at one end
in both stages of the analysis. In the first stage, we perform a quasi-static simulation
using a Newton-Raphson solver with the linearization provided in \ref{sec:appendixB}
to bring the spaghetti rod to a curvature $\kappa_{0}$. Specifically, we apply a nodal moment of 
$EI\kappa_{0}$ at the free end of the rod in $10$ load steps. 
Note that this initial pre-loading stage is necessary
because of our model's assumption that the beam be straight in the reference (unstressed)
configuration. The state achieved through the quasi-static pre-loading stage is then used as the initial condition
for a dynamic simulation starting with the release of the spaghetti rod's free end.
The physical properties and numerical parameters used in this computational experiment are
summarized in Table~\ref{tab3}. Specifically, we used the length $L$, radius $R$, and density $\rho$ 
values reported in Heisser et al. ~\cite{heisser2018controlling} and we computed the Young's modulus
$E$ of the spaghetti rods with the formula $E = \gamma^2 \rho A / I$, where $\gamma = 0.521\:m^2/s$, 
see \cite{audoly2005fragmentation}. Finally, we obtained $\sigma_c$, and $G_c$ based on the time and 
deformation shape of the rod reported in the experiments just before fracture.

\begin{figure}[h!]
	\begin{center}
		\includegraphics[width=1.0\textwidth]{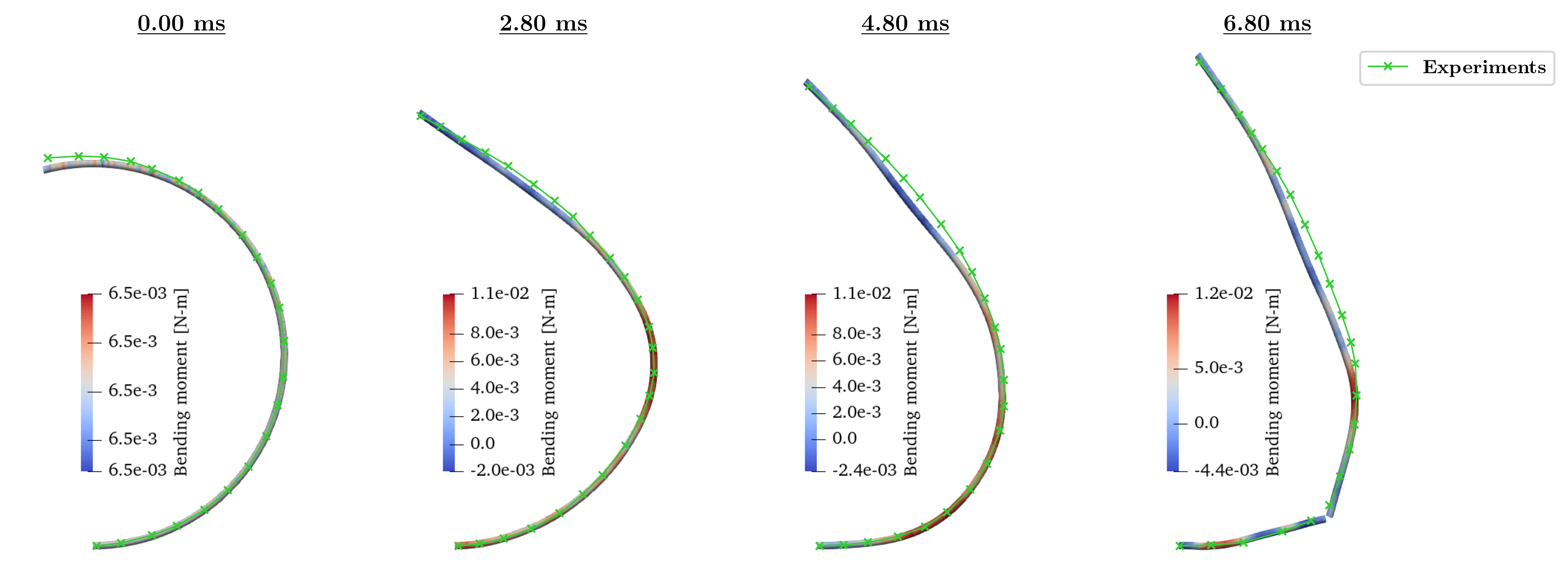}
	\end{center}
	\caption{\textit{Validation of our computational framework against fracture experiments of bent
    and released spaghetti rods~\cite{audoly2005fragmentation}.
    The pictures show the deformed spaghetti shapes obtained in our computational predictions
    overlaid to those observed experimentally (green line).
    Our DG/CZM framework captures the local curvature build-up due to the flexural waves generated
    by the release, the resulting increase of bending moment (shown in the contours),
    as well as the ultimate fracture event.
    An animation of the time evolution of this simulation is provided in the Supplementary Material.
    }}
	\label{fig12}
\end{figure}

Figure~\ref{fig12} shows contours of bending moments on the deformed spaghetti rods, superimposed
to the experimentally observed deformed shapes at different times.
Our DG/CZM framework accurately captures the fracture mechanism observed in the experiments,
including the burst of flexural waves and the curvature build-up in the proximity of the clamped end.

{\def\arraystretch{1.25}\tabcolsep=15pt
	\begin{table}[htb!]
		\begin{center}
			\begin{tabular}{ c c }
				\hline
				Property / Parameter                            & Value \\
				\hline
                Length ($L$)                                    & $0.24 \:m$\\
                Radius ($R$)                                    & $0.57\:mm$\\
                Mass density ($\rho$)                           & $1400\:kg\ m^{-3}$ \\
				Young's modulus (E)                             & $5.5\:GPa$ \\
				Critical cohesive strength ($\sigma_c$)         & $25\:MPa$ \\
				Fracture energy ($G_c$)                         & $1500\:N\ m^{-1}$ \\
				Mode-mixity parameter ($\alpha$)                & $1$ \\
				Initial curvature ($\kappa_{0}$)                 & $14.18\:rad\:s^{-1}$ \\
				Simulation time ($T$)                           & $0.01\:s$ \\
				Mesh size ($h$)                                 & $2.4\:mm$ \\
				Time step ($\Delta t$)                          & $0.1\:\mu s$ \\
				Penalty parameters ($\beta_{p,n}$ and $\beta_{t,n}$) & $10$ \\
				\hline
			\end{tabular}
			\caption{\textit{Physical properties and numerical parameters used in the bent and
            released spaghetti problem.}}
			\label{tab3}
		\end{center}
	\end{table}
}

\pagestyle{plain}

\FloatBarrier
\noindent
\section{Conclusions} \label{sec:conclusions}

We presented a computational framework to simulate the tensile and bending modes of fracture in slender beams 
subjected to large deformations.
We adopted the geometrically exact Kirchhoff beam finite element formulation by Meier et al. 
\cite{meier2015locking} to model the complex geometric nonlinearities involved in the beam deformation.
We developed a discontinuous Galerkin discretization of the beam governing equations incorporating 
jumps in position and tangent degrees of freedom.
In our framework, compatibility of nodal positions and tangents is weakly enforced before 
fracture initiation via the exchange of variationally-consistent forces and moments at the 
interfaces between adjacent elements. 
At the onset of fracture, these forces and moments transition to cohesive laws modeling 
softening of the stress-resultant forces and moments with the increasing interface separation.
We showed that incorporating discontinuities in the tangent degrees of freedom across adjacent 
elements is essential for capturing beam fracture under bending.
We conducted a series of numerical tests to verify our framework's ability to capture tensile and 
bending fracture modes in slender beams.
Finally, we applied our computational framework to reproduce experiments by Audoly and Neukirch 
\cite{audoly2005fragmentation} on the dynamic fracture behavior of dry spaghetti rods exhibiting 
large deformations.

\noindent
\section*{Acknowledgements}
This material is based upon work supported by the Air Force Office of Scientific Research under award
number FA8655-22-1-7035.
Any opinions, findings, and conclusions or recommendations expressed in this material are those of 
the author(s) and do not necessarily reflect the views of the United States Air Force.

The authors thank Dr. Sergio Turteltaub from the Delft University of Technology for providing 
insightful feedback on this research work during several conversations.

\noindent
\section*{Declaration of generative AI and AI-assisted technologies in the writing process}
During the preparation of this work the author(s) used ChatGPT in order to expedite the language 
editing process (e.g. to rewrite particularly complex sentences in a clearer way). 
After using this tool/service, the author(s) reviewed and edited the content as needed and take(s) 
full responsibility for the content of the publication.

\bibliography{bibliography}

\clearpage
\appendix
\pagestyle{plain}

\section{Inertia, internal, and external forces in the semi-discrete system of equations} \label{sec:appendixA}
This Appendix reports the expressions of the inertia, internal (bulk), internal (interface), and 
external forces discussed in Section \ref{sec:space-time}.
\begin{gather*}
	\boldsymbol{M_{ab}}\boldsymbol{\ddot{x}_{b}} = \Bigg[\int_{-1}^{1}N_{a}\rho AN_{b}\frac{L}{2}\:d\xi\Bigg]\boldsymbol{\ddot{x}_{b}}, \\
	\boldsymbol{f^{int}_{a}} = \int_{-1}^{1} \big[N_{a}^{\prime}(EA\boldsymbol{t_{1}}+EI\boldsymbol{t_{2}})+N_{a}^{\prime\prime}EI\boldsymbol{t_{3}}\big]\frac{L}{2}\:d\xi, \\
	\boldsymbol{f^{jump}_{a^{\pm}}} = \pm \sum_{n=1}^{E-1}\:\alpha_n\Bigg[\Big[N_{a}\boldsymbol{f_{DG,\:||}}\Big] + \Big[N_{a}^{\prime}\big[\langle\boldsymbol{m_{\perp}}\rangle\times\boldsymbol{t_{4}}\big]\Big] + \beta_{p}\bigg\langle\frac{EA}{h}\bigg\rangle\Big[N_{a}\boldsymbol{c_{DG,\:||}}\Big] \\
	+\: \beta_{t}\bigg\langle\frac{EI}{h}\bigg\rangle \Big[N_{a}^{\prime}\,\boldsymbol{t_5}\Big]\Bigg]\Bigg|_{s_{n}} \pm \sum_{n=1}^{E-1}\:\gamma_n \Bigg[\Big[N_{a}\boldsymbol{f_{DG,\:\perp}}\Big] + \beta_{p}\bigg\langle\frac{EA}{h}\bigg\rangle\Big[N_{a}\boldsymbol{c_{DG,\:\perp}}\Big]\Bigg]\Bigg|_{s_{n}} \\
	\pm \sum_{n=1}^{E-1}\:(1-\alpha_n)\Bigg[\Big[N_{a}\boldsymbol{f_{coh,\:||}}\Big] + \Big[N_{a}^{\prime}\boldsymbol{G_1}\boldsymbol{m_{coh,\:\perp}}\Big]\Bigg]\Bigg|_{s_{n}}, \\
	\boldsymbol{f^{ext}_{a}} = \int_{-1}^{1}\big[N_{a}\boldsymbol{\tilde{f}} + N_{a}^{\prime}(\boldsymbol{\tilde{m}_{\perp}} \times \boldsymbol{t_{4}})\big]\frac{L}{2}\:d\xi + \big[N_{a}\boldsymbol{\bar{f}}\,\big]\big|_{\partial_{N_f}\Omega_h} + \big[N_{a}^{\prime}(\boldsymbol{\bar{m}_{\perp}} \times \boldsymbol{t_{4}})\big]\big|_{\partial_{N_m}\Omega_h},
\end{gather*}
where
\begin{gather*}
	\boldsymbol{t_{1}} = \frac{\boldsymbol{r^{\prime}}}{||\boldsymbol{r^{\prime}}||}\big(||\boldsymbol{r^{\prime}}||-1\big), \\
	\boldsymbol{t_{2}} = \frac{2\boldsymbol{r^{\prime}}\big(\boldsymbol{r^{\prime}}\cdot\boldsymbol{r^{\prime\prime}}\big)^2}{||\boldsymbol{r^{\prime}}||^6} - \Bigg[\frac{\boldsymbol{r^{\prime}}\big(\boldsymbol{r^{\prime\prime}}\cdot\boldsymbol{r^{\prime\prime}}\big) + \boldsymbol{r^{\prime\prime}}\big(\boldsymbol{r^{\prime}}\cdot\boldsymbol{r^{\prime\prime}}\big)}{||\boldsymbol{r^{\prime}}||^4}\Bigg], \\
	\boldsymbol{t_{3}} = \frac{\boldsymbol{r^{\prime\prime}}}{||\boldsymbol{r^{\prime}}||^2} - \frac{\boldsymbol{r^{\prime}}\big(\boldsymbol{r^{\prime}}\cdot\boldsymbol{r^{\prime\prime}}\big)}{||\boldsymbol{r^{\prime}}||^4}, \\
	\boldsymbol{t_{4}} = \frac{\boldsymbol{r^{\prime}}}{||\boldsymbol{r^{\prime}}||^{2}}, \\
	\boldsymbol{G_1} = \frac{\boldsymbol{I}}{||\boldsymbol{r^{\prime}}||} - \frac{(\boldsymbol{r^{\prime}}\otimes\boldsymbol{r^{\prime}})}{||\boldsymbol{r^{\prime}}||^3}, \\
	\boldsymbol{t_{5}} = \boldsymbol{G_1}\jump{\boldsymbol{g_{1}}}.
\end{gather*}

\section{Linearizations of internal and external forces} \label{sec:appendixB}
This Appendix reports the expressions of the linearization of the inertia, internal (bulk), 
internal (interface), and external forces discussed in Section \ref{sec:space-time}.
\begin{gather*}
	\boldsymbol{K_{ab}^{int}} = \frac{\partial \boldsymbol{f^{int}_{a}}}{\partial \boldsymbol{x_{b}}} = \int_{-1}^{1} \Bigg[N_{a}^{\prime}\Bigg(EA\:\frac{\partial \boldsymbol{t_{1}}}{\partial \boldsymbol{x_{b}}}+EI\:\frac{\partial \boldsymbol{t_{2}}}{\partial \boldsymbol{x_{b}}}\Bigg)+N_{a}^{\prime\prime}EI\:\frac{\partial \boldsymbol{t_{3}}}{\partial \boldsymbol{x_{b}}}\Bigg]\frac{L}{2}\:d\xi, \\
	\hspace{0.1cm}
	\boldsymbol{K_{ab^{\pm}}^{jump,DG}} = \frac{\partial \boldsymbol{f^{jump,DG}_{a^{\pm}}}}{\partial \boldsymbol{x_{b}}} = \pm \sum_{n=1}^{E-1} \hspace{0.1cm} \Bigg[N_{a}EA\frac{\partial \langle\boldsymbol{t_{1}}\rangle}{\partial \boldsymbol{x_{b}}}\Bigg]\Bigg|_{s_{n}} \pm \sum_{n=1}^{E-1}
	\Bigg[N_{a}EI\frac{\partial \langle\boldsymbol{t_{6}}\rangle}{\partial \boldsymbol{x_{b}}}\Bigg]\Bigg|_{s_{n}} \\
	\hspace{0.1cm}
	\pm \sum_{n=1}^{E-1} \Bigg[-N_{a}\Bigg(\boldsymbol{S(\tilde{m}_{\perp})}\frac{\partial \langle\boldsymbol{t_{4}}\rangle}{\partial \boldsymbol{x_{b}}}\Bigg)\Bigg]\Bigg|_{s_{n}}
	\pm \sum_{n=1}^{E-1} \Bigg[N_{a}^{\prime}\Bigg(\boldsymbol{S(\langle m_{\perp}\rangle)}\frac{\partial \boldsymbol{t_{4}}}{\partial \boldsymbol{x_{b}}}\Bigg)\Bigg]\Bigg|_{s_{n}} \\
	\hspace{0.1cm}
	\pm \sum_{n=1}^{E-1} \Bigg[-N_{a}^{\prime}\Bigg(\boldsymbol{S(t_{4})}\frac{\partial \langle\boldsymbol{m_{\perp}}\rangle}{\partial \boldsymbol{x_{b}}}\Bigg)\Bigg]\Bigg|_{s_{n}} 
	\pm \sum_{n=1}^{E-1} \beta_{p}\bigg\langle\frac{EA}{h}\bigg\rangle \boldsymbol{I}\Big[N_{a}N_{b}\Big]\Big|_{s_{n}} \\
	\hspace{0.1cm}
	\pm \sum_{n=1}^{E-1} \beta_{t}\bigg\langle\frac{EI}{h}\bigg\rangle
	\Bigg[N_{a}^{\prime}\,\frac{\partial\boldsymbol{t_{5}}}{\partial \boldsymbol{x_{b}}}\Bigg]\Bigg|_{s_{n}}, \\
	\hspace{0.1cm}
	\boldsymbol{K_{ab}^{ext}} = \frac{\partial \boldsymbol{f^{ext}_{a}}}{\partial \boldsymbol{x_{b}}} =
	\int_{-1}^{1}\Bigg[N_{a}^{\prime}\Bigg(\boldsymbol{S(\tilde{m}_{\perp})}\:\frac{\partial \boldsymbol{t_{4}}}{\partial \boldsymbol{x_{b}}}\Bigg)\Bigg]\frac{L}{2}\:d\xi +
	\Bigg[N_{a}^{\prime}\Bigg(\boldsymbol{S(\bar{m}_{\perp})}\:\frac{\partial \boldsymbol{t_{4}}}{\partial \boldsymbol{x_{b}}}\Bigg)\Bigg]\Bigg|_{\partial_{N_m}\Omega_h},
\end{gather*}
where $\boldsymbol{S}(\cdot)$ is a skew-symmetric matrix such that $\boldsymbol{S(a)b} 
= \boldsymbol{a} \times \boldsymbol{b}$ and
\begin{gather*}
	\frac{\partial \boldsymbol{t_{1}}}{\partial \boldsymbol{x_{b}}} = \Bigg[\frac{(||\boldsymbol{r^{\prime}}|| -
			1)}{||\boldsymbol{r^{\prime}}||}\boldsymbol{I} +
		\frac{(\boldsymbol{r^{\prime}}\otimes\boldsymbol{r^{\prime}})}{||\boldsymbol{r^{\prime}}||^3}\Bigg]N_{b}^{\prime},
	\\ \frac{\partial \boldsymbol{t_{2}}}{\partial \boldsymbol{x_{b}}} =
	\Bigg[\Bigg[\frac{2\big(\boldsymbol{r^{\prime}}\cdot\boldsymbol{r^{\prime\prime}}\big)^2}{||\boldsymbol{r^{\prime}}||^6}
			-
			\frac{\big(\boldsymbol{r^{\prime\prime}}\cdot\boldsymbol{r^{\prime\prime}}\big)}{||\boldsymbol{r^{\prime}}||^4}\Bigg]\boldsymbol{I}
		+
		\Bigg[\frac{-12\big(\boldsymbol{r^{\prime}}\cdot\boldsymbol{r^{\prime\prime}}\big)^2}{||\boldsymbol{r^{\prime}}||^8}
			+
			\frac{4\big(\boldsymbol{r^{\prime\prime}}\cdot\boldsymbol{r^{\prime\prime}}\big)}{||\boldsymbol{r^{\prime}}||^6}\Bigg](\boldsymbol{r^{\prime}}\otimes\boldsymbol{r^{\prime}})
		\\ +
		\frac{4\big(\boldsymbol{r^{\prime}}\cdot\boldsymbol{r^{\prime\prime}}\big)}{||\boldsymbol{r^{\prime}}||^6}(\boldsymbol{r^{\prime}}\otimes\boldsymbol{r^{\prime\prime}})
		+
		\frac{4\big(\boldsymbol{r^{\prime}}\cdot\boldsymbol{r^{\prime\prime}}\big)}{||\boldsymbol{r^{\prime}}||^6}(\boldsymbol{r^{\prime\prime}}\otimes\boldsymbol{r^{\prime}})
		-
		\frac{(\boldsymbol{r^{\prime\prime}}\otimes\boldsymbol{r^{\prime\prime}})}{||\boldsymbol{r^{\prime}}||^4}\Bigg]N_{b}^{\prime}
	\\ +
	\Bigg[\frac{-\big(\boldsymbol{r^{\prime}}\cdot\boldsymbol{r^{\prime\prime}}\big)}{||\boldsymbol{r^{\prime}}||^4}\boldsymbol{I}
		+
		\frac{4\big(\boldsymbol{r^{\prime}}\cdot\boldsymbol{r^{\prime\prime}}\big)}{||\boldsymbol{r^{\prime}}||^6}(\boldsymbol{r^{\prime}}\otimes\boldsymbol{r^{\prime}})
		-
		\frac{2(\boldsymbol{r^{\prime}}\otimes\boldsymbol{r^{\prime\prime}})}{||\boldsymbol{r^{\prime}}||^4}
		-
		\frac{(\boldsymbol{r^{\prime\prime}}\otimes\boldsymbol{r^{\prime}})}{||\boldsymbol{r^{\prime}}||^4}\Bigg]N_{b}^{\prime\prime},
	\\ \frac{\partial \boldsymbol{t_{3}}}{\partial \boldsymbol{x_{b}}} =
		\Bigg[\frac{-\big(\boldsymbol{r^{\prime}}\cdot\boldsymbol{r^{\prime\prime}}\big)}{||\boldsymbol{r^{\prime}}||^4}\boldsymbol{I}
			+
			\frac{4\big(\boldsymbol{r^{\prime}}\cdot\boldsymbol{r^{\prime\prime}}\big)}{||\boldsymbol{r^{\prime}}||^6}(\boldsymbol{r^{\prime}}\otimes\boldsymbol{r^{\prime}})
			-
			\frac{2(\boldsymbol{r^{\prime\prime}}\otimes\boldsymbol{r^{\prime}})}{||\boldsymbol{r^{\prime}}||^4}
			-
			\frac{(\boldsymbol{r^{\prime}}\otimes\boldsymbol{r^{\prime\prime}})}{||\boldsymbol{r^{\prime}}||^4}\Bigg]N_{b}^{\prime}
		\\ + \Bigg[\frac{\boldsymbol{I}}{||\boldsymbol{r^{\prime}}||^2} -
			\frac{(\boldsymbol{r^{\prime}}\otimes\boldsymbol{r^{\prime}})}{||\boldsymbol{r^{\prime}}||^4}\Bigg]N_{b}^{\prime\prime},
		\\ \frac{\partial \boldsymbol{t_{4}}}{\partial \boldsymbol{x_{b}}} = \Bigg[\frac{\boldsymbol{I}}{||\boldsymbol{r^{\prime}}||^2} -
			\frac{2(\boldsymbol{r^{\prime}}\otimes\boldsymbol{r^{\prime}})}{||\boldsymbol{r^{\prime}}||^4}\Bigg]N_{b}^{\prime},
		\\ \frac{\partial\boldsymbol{m_{\perp}}}{\partial \boldsymbol{x_{b}}} = 
		EI\Bigg[-\boldsymbol{S(\boldsymbol{r^{\prime\prime}})}\frac{\partial\boldsymbol{t_{4}}}{\partial \boldsymbol{x_{b}}}
		+ \boldsymbol{S(\boldsymbol{t_{4}})}N_{b}^{\prime\prime}\Bigg],
		\\ \frac{\partial \boldsymbol{t_{5}}}{\partial \boldsymbol{x_{b}}} = \Bigg[\boldsymbol{G_1}\boldsymbol{G_1} 
		- \frac{(\jump{\boldsymbol{g_{1}}}\otimes\boldsymbol{r^{\prime}})}{||\boldsymbol{r^{\prime}}||^3}
		- \frac{(\boldsymbol{r^{\prime}}\otimes\jump{\boldsymbol{g_{1}}})}{||\boldsymbol{r^{\prime}}||^3}
		- \frac{(\jump{\boldsymbol{g_{1}}}\cdot\boldsymbol{r^{\prime}})}{||\boldsymbol{r^{\prime}}||^3}\boldsymbol{I}
		+ \frac{3(\jump{\boldsymbol{g_{1}}}\cdot\boldsymbol{r^{\prime}})(\boldsymbol{r^{\prime}}\otimes\boldsymbol{r^{\prime}})}{||\boldsymbol{r^{\prime}}||^5}\Bigg]N_{b}^{\prime},
		\\ \boldsymbol{t_{6}} = \boldsymbol{t_{4}} \times \boldsymbol{\kappa^{\prime}} = \frac{2\boldsymbol{r^{\prime\prime}}\big(\boldsymbol{r^{\prime}}\cdot\boldsymbol{r^{\prime\prime}}\big)}{||\boldsymbol{r^{\prime}}||^4} + \frac{\boldsymbol{r^{\prime}}\big(\boldsymbol{r^{\prime}}\cdot\boldsymbol{r^{\prime\prime\prime}}\big)}{||\boldsymbol{r^{\prime}}||^4} - \Bigg[\frac{2\boldsymbol{r^{\prime}}\big(\boldsymbol{r^{\prime}}\cdot\boldsymbol{r^{\prime\prime}}\big)^2}{||\boldsymbol{r^{\prime}}||^6} + \frac{\boldsymbol{r^{\prime\prime\prime}}}{||\boldsymbol{r^{\prime}}||^2}\Bigg], \\
	\frac{\partial \boldsymbol{t_{6}}}{\partial \boldsymbol{x_{b}}} =
	\Bigg[\Bigg[\frac{-2\big(\boldsymbol{r^{\prime}}\cdot\boldsymbol{r^{\prime\prime}}\big)^2}{||\boldsymbol{r^{\prime}}||^6}
			+
			\frac{\big(\boldsymbol{r^{\prime}}\cdot\boldsymbol{r^{\prime\prime\prime}}\big)}{||\boldsymbol{r^{\prime}}||^4}\Bigg]\boldsymbol{I}
		+
		\Bigg[\frac{12\big(\boldsymbol{r^{\prime}}\cdot\boldsymbol{r^{\prime\prime}}\big)^2}{||\boldsymbol{r^{\prime}}||^8}
			-
			\frac{4\big(\boldsymbol{r^{\prime}}\cdot\boldsymbol{r^{\prime\prime\prime}}\big)}{||\boldsymbol{r^{\prime}}||^6}\Bigg](\boldsymbol{r^{\prime}}\otimes\boldsymbol{r^{\prime}})
		\\ -
		\frac{4\big(\boldsymbol{r^{\prime}}\cdot\boldsymbol{r^{\prime\prime}}\big)}{||\boldsymbol{r^{\prime}}||^6}(\boldsymbol{r^{\prime}}\otimes\boldsymbol{r^{\prime\prime}})
		-
		\frac{8\big(\boldsymbol{r^{\prime}}\cdot\boldsymbol{r^{\prime\prime}}\big)}{||\boldsymbol{r^{\prime}}||^6}(\boldsymbol{r^{\prime\prime}}\otimes\boldsymbol{r^{\prime}})
		+
		\frac{2(\boldsymbol{r^{\prime\prime}}\otimes\boldsymbol{r^{\prime\prime}})}{||\boldsymbol{r^{\prime}}||^4}
		+
		\frac{(\boldsymbol{r^{\prime}}\otimes\boldsymbol{r^{\prime\prime\prime}})}{||\boldsymbol{r^{\prime}}||^4}
		+
		\frac{2(\boldsymbol{r^{\prime\prime\prime}}\otimes\boldsymbol{r^{\prime}})}{||\boldsymbol{r^{\prime}}||^4}\Bigg]N_{b}^{\prime}
	\\ +
	\Bigg[\frac{2\big(\boldsymbol{r^{\prime}}\cdot\boldsymbol{r^{\prime\prime}}\big)}{||\boldsymbol{r^{\prime}}||^4}\boldsymbol{I}
		-
		\frac{4\big(\boldsymbol{r^{\prime}}\cdot\boldsymbol{r^{\prime\prime}}\big)}{||\boldsymbol{r^{\prime}}||^6}(\boldsymbol{r^{\prime}}\otimes\boldsymbol{r^{\prime}})
		+
		\frac{2(\boldsymbol{r^{\prime\prime}}\otimes\boldsymbol{r^{\prime}})}{||\boldsymbol{r^{\prime}}||^4}\Bigg]N_{b}^{\prime\prime}
	+ \Bigg[\frac{-\boldsymbol{I}}{||\boldsymbol{r^{\prime}}||^2} +
		\frac{(\boldsymbol{r^{\prime}}\otimes\boldsymbol{r^{\prime}})}{||\boldsymbol{r^{\prime}}||^4}\Bigg]N_{b}^{\prime\prime\prime}.
\end{gather*}

\end{document}